\documentclass[pdftex,11pt,a4paper,parskip=half,headsepline]{scrartcl}

\usepackage[english]{babel}
\usepackage[a4paper,left=3cm,right=3cm,bottom=3cm,top=3cm]{geometry}
\usepackage[automark]{scrlayer-scrpage}
\usepackage[utf8]{inputenc}
\usepackage{latexsym}
\usepackage[T1]{fontenc}
\usepackage{amsmath}
\usepackage{amssymb}
\usepackage{amsfonts}
\usepackage{amstext}
\usepackage{dsfont}
\usepackage{amsthm}
\usepackage{extarrows}
\usepackage{graphicx}
\usepackage{xcolor}
\usepackage{longtable}
\usepackage{enumitem}
\usepackage[longtable]{multirow} 
\usepackage{booktabs} 
\usepackage{tabularx}
\usepackage{url}
\usepackage{algorithm}
\usepackage{algpseudocode}
\usepackage{csquotes}

\usepackage[backend=biber, bibstyle=apa, maxcitenames=2, citestyle=apa, isbn=false, doi=true, url=false, eprint=true]{biblatex}
\AtEveryBibitem{\clearfield{month}}
\AtEveryCitekey{\clearfield{month}}
\AtEveryBibitem{\clearfield{day}}
\AtEveryCitekey{\clearfield{day}}
\AtEveryBibitem{\clearfield{note}}
\AtEveryCitekey{\clearfield{note}}
\NewBibliographyString{toappear}
\DefineBibliographyStrings{english}{%
  toappear = {to appear in:},
  version = {R package version},
} 
\usepackage[colorlinks=true,linkcolor=black,citecolor=black,filecolor=black,urlcolor=black]{hyperref}
\addto\extrasenglish{

}

\title{Resampling-Based Inference for Incomplete Paired Functional Data}

\subtitle{}

\author{Marl\'{e}ne Baumeister$^{\dagger,1}$, Lubna Amro$^{1}$,\\
	 \L ukasz Smaga$^{3}$, Markus Pauly$^{1,2}$}

\makeatletter

\clearpairofpagestyles
\ihead{\headmark}
\automark{section}
\ohead{Functional Matched Pairs with Missing Values}
\newcolumntype{R}{>{\raggedleft\arraybackslash}X}

\newtheorem{satz}{Theorem}[]

\newtheorem{prop}[satz]{Proposition}
\newtheorem{ass}[satz]{Assumption}

\newtheorem{cor}[satz]{Corollary}

\newcommand{\br}{\mathbb{R}}
\newcommand{\bn}{\mathbb{N}}

\DeclareMathOperator{\prob}{Pr}
\DeclareMathOperator{\E}{E}
\DeclareMathOperator{\D}{D} 
\DeclareMathOperator{\var}{Var}
\DeclareMathOperator{\cov}{Cov}

\DeclareMathOperator{\GP}{GP}
\DeclareMathOperator{\SP}{SP}

\newcommand{\vertiii}[1]{{\left\vert\kern-0.25ex\left\vert\kern-0.25ex\left\vert #1 
		\right\vert\kern-0.25ex\right\vert\kern-0.25ex\right\vert}} 

\definecolor{bblue}{HTML}{28539E}
\definecolor{turquoise}{HTML}{47C2D6}
\definecolor{pink}{HTML}{E39AB7}
\definecolor{lgreen}{HTML}{94e017 }
\definecolor{orange}{HTML}{e88909}
\definecolor{sagegreen}{RGB}{143, 188, 143}     

\begin{document}

\thispagestyle{empty}
\maketitle

\renewcommand*{\thefootnote}{\fnsymbol{footnote}}
\footnotetext[2]{Corresponding author: Marl\'{e}ne Baumeister, \textsf{e-mail: baumeister@statistik.tu-dortmund.de}}
\renewcommand*{\thefootnote}{\arabic{footnote}}
\footnotetext[1]{Department of Statistics, TU Dortmund University, Germany}
\footnotetext[2]{Research Center Trustworthy Data Science and Security, UA Ruhr, Germany}
\footnotetext[3]{Faculty of Mathematics and Computer Science, Adam Mickiewicz University in Pozna{\'n}, Poland}

\section*{Abstract}
We consider testing equality of mean functions for paired functional data, where one entire function is missing.
To use all available information from the paired and unpaired observations, we propose globalized test statistics that are based on a weighted sum of point-wise paired-$t$ and Welch-$t$-type statistics. 
By the use of empirical process theory, we derive the asymptotic limit distributions of the resulting test statistics under the null hypothesis and show consistency against fixed alternatives. 
As they are non-pivotal and contain complex dependencies from the paired and incomplete observations, a permutation and parametric bootstrap procedures is proposed to compute critical values. 
The work includes a theoretical investigation of the asymptotic behaviour of the two resampling approaches and state regularity conditions under which the corresponding resampling tests are valid. 
Simulation studies show that the proposed methods provide accurate type I error control and competitive power for most settings under study.
An illustrative analysis of functional emission data demonstrates the applicability of the methods.

\textbf{Keywords:} 
Matched pairs, Missing data, Parametric Bootstrap, Permutation test, Two-Sample Testing.

\section[Introduction]{Introduction and Motivation}\label{sec:intro}
Functional data analysis (\textsc{fda}) offers statistical methods for observations that can be represented as curves or functions. 
This type of data arises increasingly often due to modern measurement technology that record signals densely over time or space \parencite{gertheiss_functional_2024}.
\textsc{fda} methods allow the analysis of such data while accounting for the correlation structure between sampled observations.
Therefore, \textsc{fda} has become relevant in biostatistics and many other fields in which high-dimensional repeated measurements are collected \parencite{ullah_applications_2013,ramsay2002applied,gertheiss_functional_2024}.
For these reasons, \textcite{ullah_applications_2013} have pointed out the need for methodological research on \textsc{fda} and \textcite{gertheiss_functional_2024} still have emphasized that a decade later.
Despite substantial progress, the handling of missing data remains challenging.
A large part of the literature considers partially observed functional data,
where individual trajectories are available only on subsets 
\parencite[e.g.][]{kraus_components_2015, kraus_inferential_2019, liebl_partially_2019, kneip_optimal_2020}.
However, there is less work on statistical problems in which an entire function is missing. 
This problem is particularly relevant for paired functional data.

In this paper, we develop testing procedures for incomplete paired functional data. 
Our approach adapts the idea of \textcite{amro_permuting_2017} to functional data. 
Rather than imputing the missing functions, as in the approach of 
\textcite{jang_bayesian_2021}, we consider a weighted sum of test statistics constructed from the complete pairs and from the unpaired observations.
This allows us to include all observed functions in the analysis.
Two resampling approaches, a permutation procedure and a parametric bootstrap, were used to compute the critical values.
The proposed methods are developed under a missing completely at random (\textsc{mcar}) mechanism and works without any distributional or homoscedasticity assumption.
This way, we extend the complete-pairs procedure of \textcite{smaga_note_2020} to be applicable in the situation of missing functions.
We prove the asymptotic validity of the parametric bootstrap procedure without requiring specific assumptions on the covariance functions. 
We further show that, contrary to the complete paired setting  \parencite{smaga_note_2020}, the permutation procedure is not valid in general, but only under certain conditions. 
To investigate the finite-sample properties of both procedures, we conducted an extensive simulation.
We thereby also compare them with the bootstrap-based approach of \textcite{crainiceanu2012bootstrap} and with the method of \textcite{smaga_note_2020}, applied both directly and in combination with the imputation approach of \textcite{jang_bayesian_2021}. 
The results indicate that both approaches generally control the type I error well under the investigated settings and have comparable power. 
The advantages are particularly present for small sample sizes and for larger proportions of incomplete pairs.

Our main contribution is the development of general inference procedures for incomplete functional matched pairs that
\begin{enumerate}
	\item use the entire observed functional data,
	\item work well on small samples and high proportions of missing functions, 
	\item and do not require strong distributional assumption such as Gaussianity or equality of the covariance operators.
\end{enumerate}

The paper is structured as follows. 
In Section~\ref{sec:setup}, we introduce the statistical model for incomplete paired functional data, including its assumptions, and the hypothesis of interest.
The estimators, test statistics, and their asymptotic properties are stated in Section~\ref{sec:methods}.
Section~\ref{sec:resampling} introduces the permutation and bootstrap procedures and explains the conditions required for their validity, while Section~\ref{sec:simulation} describes the simulation study design and compares the proposed methods with existing alternatives. 
In Section~\ref{sec:real_data_example}, we apply the methods to a real data example. 
Conclusions and a discussion are presented in Section~\ref{sec:end}.
All proofs and additional theoretical results are given in Section~\ref{sec:proofs}. 
Additional simulation results are provided in the Supplementary Material \parencite{baumeister_supplementary_2026}.

\section[Model]{A Model of Matched Pairs for Functional Data}\label{sec:setup}

Let $(\Omega, \mathcal{A}, P)$ be a probability space, $(\mathcal T,\zeta)$ be a compact pseudometric space (e.g. $\mathcal T=[0,1]$ equipped with the Euclidean metric) and let $C^p(\mathcal{T})=\{f:\,\mathcal{T}\to\br^p\,|\,f\;\text{continuous}\}$ denote the space of $p$-dimensional continuous functions on $\mathcal{T}$. 
We consider a bivariate stochastic process
 $\{\mathbf{X}(t)=(X_1(t),X_2(t))^{\top}:t\in\mathcal{T}\}$,  $\mathbf{X}:\,\Omega\to C^2(\mathcal{T}),\;\boldsymbol{\omega}\mapsto\mathbf{X}_t(\boldsymbol{\omega})$,
with mean function $\boldsymbol{\eta}:\mathcal{T}\to\br^2$ and covariance function $\boldsymbol{\Gamma}:\mathcal{T}^2\to\br^{2\times 2}$. 
Shortly, we write $\mathbf{X}\sim \SP_2(\boldsymbol{\eta},\boldsymbol{\Sigma})$ or $\mathbf{Z}\sim \GP_d(\E(\mathbf{Z}),\cov(\mathbf{Z}))$ for a $d$-dimensional Gaussian process with mean function $\E(\mathbf{Z})$ and covariance function $\cov(\mathbf{Z})$.
We then consider a functional matched pairs design given by $n\in\bn$ independent realizations of $\mathbf{X}$:
\begin{align}\label{eq:data}
	\mathbf{x}_1,\dots,\mathbf{x}_n: \mathcal{T}\to\br^2,\quad \mathbf{x}_j(t) = \left(x_{1j}(t), x_{2j}(t)\right)^\top,\quad j\in\{1,\dots,n\}.
\end{align}
For some subjects, one of the two component functions may be completely unobserved for every $t\in\mathcal{T}$.
To handle this, for every $t\in\mathcal{T}$, we divide the observations in Equation~(\ref{eq:data}) into $n_1$ complete pairs, $n_2$ incomplete pairs for which only the first component is observed, and $n_3$ incomplete pairs for which only the second component is observed ($n_2+n_3=n-n_1$):
\begin{align}\label{eq:missingdata}
	\underbrace{
		\begin{pmatrix}
			x_{11}^{(c)}(t)\\
			x_{21}^{(c)}(t)
		\end{pmatrix},
		\dots,
		\begin{pmatrix}
			x_{1n_1}^{(c)}(t)\\
			x_{2n_1}^{(c)}(t)
		\end{pmatrix} 
	}_{ \mathbf{x}^{(c)}(t)}
	,
	\underbrace{
		\begin{pmatrix}
			x_{11}^{(m)}(t)\\
			---
		\end{pmatrix},
		\dots,
		\begin{pmatrix}
			x_{1n_2}^{(m)}(t)\\
			---
		\end{pmatrix},
		\begin{pmatrix}
			---\\
			x_{21}^{(m)}(t)
		\end{pmatrix},
		\dots,
		\begin{pmatrix}
			---\\
			x_{2n_3}^{(m)}(t)
		\end{pmatrix}
	}_ {\mathbf{x}^{(m)}(t)}.
\end{align}
As the observations are independent realizations of the stochastic process $\mathbf{X}$, the complete observations
$\mathbf{x}^{(c)}_j=\bigl(x_{1j}^{(c)},x_{2j}^{(c)}\bigr)^\top,
j\in\{1,\dots,n_1\},$
are independent identically distributed (i.i.d.) with mean function $\boldsymbol{\eta}=(\eta_1,\eta_2)^\top$ and covariance function $ \boldsymbol{\Gamma}$ given by
\begin{align}\label{eq:covariance}
	\boldsymbol{\Gamma}(s,t)=
	\begin{pmatrix}
		\gamma_1(s,t) & \gamma_{12}(s,t)\\
		\gamma_{12}(s,t) & \gamma_2(s,t)
	\end{pmatrix},
	\quad s,t\in\mathcal{T}.
\end{align}
We assume that the missingness mechanism is missing completely at random (\textsc{mcar}). 
Consequently, the distribution of the observed functions does not depend on the missingness pattern. 
In particular, the observed first-component functions in the incomplete sample have the same marginal distribution as $X_1$, and the observed second-component functions have the same marginal distribution as $X_2$.
The hypothesis of interest is equality of the two mean functions:
\begin{align}\label{eq:hypothesis}
	\mathcal{H}_0:\eta_1(t)=\eta_2(t)
    \text{ for all }t\in\mathcal{T},\quad
	\mathcal{H}_1:  \eta_1(t)\neq\eta_2(t)
    \text{ for at least one }t\in\mathcal{T}.
\end{align}
We write $\stackrel{d}{\longrightarrow}$ for convergence in distribution in the sense of \textcite{van_der_vaart_weak_2023} and $\stackrel{p}{\longrightarrow}$ for convergence in probability.
In order to get inference within this model, the following assumptions are made:
\begin{ass}\label{thm:assumptions}
	Consider the data as in Equation (\ref{eq:data}) and the missing functions as in Equation (\ref{eq:missingdata}).
	Let $\mathbf{v}_{j}(t)=\mathbf{x}_{j}(t)-\boldsymbol{\eta}(t)$, $j\in\{1,\dots,n\}$, $t\in\mathcal{T}$ denote the subject-effect functions.
	\begin{enumerate}[label=(A\arabic*)]
		\item\label{ass:A1} The subject-effect functions $\mathbf{v}_j(t)$, $j\in\{1,\dots,n\}$, are i.i.d. realizations of a stochastic process $\{\mathbf{V}(t): t\in\mathcal{T}\}$, $\mathbf{V}:\,\Omega\to C^2(\mathcal{T})$ with zero mean function and covariance function $\boldsymbol{\Gamma}$ for every $t\in\mathcal{T}$.
		\item\label{ass:A2} There exist $t^*\in\mathcal{T}$ and $J\ge 2$ such that $\E[\|\mathbf{v}_1(t^*)\|_2^J]<\infty$, where here and throughout $\|\cdot\|_2$ denotes the Euclidean norm of a vector of numbers.
		\item\label{ass:A3} The pointwise covariance matrix $\boldsymbol{\Gamma}(t,t)$ is positive definite for all $t\in\mathcal{T}$.
		\item\label{ass:A4} The complete and incomplete parts 
        are asymptotically non-negligible, i.e., $n_i/(n_2+n_3)\stackrel{}{\longrightarrow}\kappa_i>0$ for $\min\{n_1,n_2,n_3\}\to \infty$, $i\in\{1,2,3\}$ almost surely.
		\item\label{ass:A5} There exists a $C\ge0$ for all $s,t\in\mathcal{T}$ such that $\E[\|\mathbf{v}_1(t)-\mathbf{v}_1(s)\|_2^J]\le C\cdot\zeta(s,t)^J$, and there is a $\nu>0$ such that
		\begin{align}\label{eq:packingnumber}
			\int_0^{\nu}\D(\epsilon, \zeta)^{2/J}d\epsilon<\infty.
		\end{align}
		Here and throughout, $J$ is chosen as in \ref{ass:A2} and $\D(\epsilon, \zeta)$ denotes the packing number, see \textcite[][Def. 2.2.3]{van_der_vaart_weak_2023}.
	\end{enumerate}
\end{ass}
Assumptions~\ref{ass:A1}-\ref{ass:A5} are empirical-process conditions used to establish weak convergence in spaces of continuous functions. 
This follows \textcite{dette_functional_2020,munko_multiple_2025} who consider stochastic processes valued in spaces of continuous functions to apply empirical process theory \parencite{van_der_vaart_weak_2023}. 
The conditions are similar to the assumptions of \textcite{munko_multiple_2025}.
Assumption \ref{ass:A4} ensures that each of the three subsamples contributes asymptotically to the test statistic, compare \textcite{amro_permuting_2017}.
The moment and packing conditions \ref{ass:A2} and \ref{ass:A5} are related to Theorem 2.2.4 in \textcite{van_der_vaart_weak_2023}.
Unless stated otherwise, all asymptotic statements are understood as $\min\{n_1,n_2,n_3\}\to\infty$.

\section{Statistical Methodology}\label{sec:methods}
To test the hypothesis in Equation~(\ref{eq:hypothesis}), we construct point-wise $t$-type statistics and subsequently globalize them by a supremum or an integration to obtain functional test statistics. 
Our approach extends the functional tests of \textcite{smaga_note_2020,munko_multiple_2025} to incomplete paired functional data.
We then define the empirical mean function $\boldsymbol{\hat{\eta}}(t)=(\hat\eta_1(t),\hat\eta_2(t))^{\top}:=1/n\sum_{j=1}^n\mathbf{x}_{j}(t)$ and  the empirical covariance function $\boldsymbol{\hat\Gamma}:\mathcal{T}^2\to\br^{2\times 2}$ 
by
\begin{equation}\label{eq:empcovariance}
	\boldsymbol{\hat\Gamma}(s,t)=\begin{pmatrix}
		\hat\gamma_1(s,t) & \hat\gamma_{12}(s,t)\\
		\hat\gamma_{12}(s,t) & \hat\gamma_2(s,t)
	\end{pmatrix}:=\frac{1}{n-1}\sum_{j=1}^n\left(\mathbf{x}_j(s)-\boldsymbol{\hat{\eta}}(s)\right)\left(\mathbf{x}_j(t)-\boldsymbol{\hat{\eta}}(t)\right)^\top, \ s,t\in\mathcal{T}.
\end{equation}
Under Assumption \ref{ass:A1}, the weak law of large numbers implies point-wise consistency of $\boldsymbol{\hat{\eta}}(t)$ and $\boldsymbol{\hat\Gamma}(s,t)$ for every $s,t\in\mathcal{T}$. 
Moreover, Lemma~6 in \textcite{munko_multiple_2025} ensures that $\boldsymbol{\hat\Gamma}$ is even uniformly consistent on $\{(t,t)\in\mathcal{T}^2\,|\,t\in\mathcal{T}\}$ under Assumptions \ref{ass:A1}, \ref{ass:A2}, and \ref{ass:A5}.
We first consider the complete-data setting as this serves as a building block. 
Therefore, let $n=n_1$. 
For this situation, \textcite{smaga_note_2020} proposed a functional analogue of the classical paired $t$-test by using the integral and the supremum to globalize the point-wise $t$-type statistics. 
To adopt this idea, we consider the point-wise test statistic:
\begin{align}
	\label{eq:paired_pointwise_statistics}
	T_{n_1}\left(\mathbf{x}^{(c)}, \boldsymbol{\hat\Gamma}^{(c)}\right)(t) := \sqrt{n_1}\,\frac{\hat{\eta}_1^{(c)}(t)- \hat{\eta}_2^{(c)}(t)}{\sqrt{\hat{\gamma}^{(c)}_{1-2}(t,t)}}, \  t \in \mathcal{T},
\end{align}
where $\boldsymbol{\hat{\eta}}^{(c)}(t)=(\hat\eta_1^{(c)}(t), \hat\eta_2^{(c)}(t))^\top,\,t\in\mathcal{T}$, is the group-wise sample mean function on the complete data $\mathbf{x}^{(c)}$ and 
\begin{align}
	\label{eq:gamma_diff}
	\left(\hat{\gamma}_{1-2}^{(c)}(t,t)\right)_{t\in\mathcal{T}}  := \left(\hat{\gamma}^{(c)}_1(t,t) - 2\hat{\gamma}^{(c)}_{12}(t,t) + \hat{\gamma}^{(c)}_2(t,t)\right)_{t\in\mathcal{T}}
\end{align}
is the empirical covariance function of the paired differences and therefore a consistent estimator of the variance $\gamma_{1-2}^{(c)}=\gamma_1^{(c)}-2\gamma_{12}^{(c)}+\gamma_2^{(c)}$ of the limiting stochastic process of $(\hat{\eta}_1^{(c)}(t)-\hat{\eta}_2^{(c)}(t))_{t\in\mathcal{T}}$.

We now return to the incomplete paired functional data model introduced in Equation~(\ref{eq:missingdata}) in Section~\ref{sec:setup}. 
The incomplete observations provide additional information about the mean functions. 
To incorporate this information, we follow the idea of  \textcite{amro_permuting_2017} and treat the observations available from the first and second components as two independent functional samples.
If there were only incomplete pairs available, i.e. $n = n_2 + n_3$, a  natural point-wise statistic would be a Welch-type statistic given by
\begin{align}
	\label{eq:missing_pointwise_statistics}
	T_{n_2+n_3}\left(\mathbf{x}^{(m)}, \boldsymbol{\hat\Gamma}_0^{(m)} \right)(t) := \frac{\hat{\eta}_1^{(m)}(t)- \hat{\eta}_2^{(m)}(t)}{\sqrt{\frac{\hat\gamma_1^{(m)}(t,t)}{n_2}+\frac{\hat\gamma_2^{(m)}(t,t)}{n_3}}}, \quad t\in \mathcal{T}.
\end{align}
Here, $\boldsymbol{\hat{\eta}}^{(m)}=(\hat\eta_1^{(m)}, \hat\eta_2^{(m)})^\top$ is the group-wise sample mean function on the incomplete data $\mathbf{x}^{(m)}$, where 
$\hat{\eta}_1^{(m)}(t) =1/n_2 \sum_{j=1}^{n_2} x_{1j}^{(m)}(t)$ 
and 
$\hat{\eta}_2^{(m)}(t) = 1/n_3 \sum_{j=1}^{n_3} x_{2j}^{(m)}(t)$.
Moreover, $\hat\gamma_1^{(m)}(t,t)$ and $\hat\gamma_2^{(m)}(t,t)$ are the 
corresponding group-wise empirical variance functions defined on the first respective second component of the observed missing data pairs $\mathbf{x}^{(m)}$, compare Equation~(\ref{eq:empcovariance}). 
We write
\begin{align*}
	\boldsymbol{\hat{\Gamma}}_0^{(m)}(s,t)
	=\begin{pmatrix}
		\frac{n_3}{n_2+n_3}\hat\gamma_1^{(m)}(s,t) & 0\\
		0 & \frac{n_2}{n_2+n_3}\hat\gamma_2^{(m)}(s,t)
	\end{pmatrix}
\end{align*}
for every $s,t\in\mathcal T$ to highlight the unbalanced sample sizes and independence of the two groups of incomplete observations.
To use all available data, we follow the idea of \textcite{amro_permuting_2017} and combine the squared paired $t$- and Welch-type statistics and define
\begin{equation} \label{eq:pointwise_weighted}
	K_n(\mathbf{x},t)=\sqrt{a_n}\,T^2_{n_1}\left(\mathbf{x}^{(c)}, \boldsymbol{\hat\Gamma}^{(c)}\right)(t) + \sqrt{1-a_n} T^2_{n_2+n_3}\left(\mathbf{x}^{(m)}, \boldsymbol{\hat\Gamma}_{0}^{(m)}\right)(t),\  t\in\mathcal{T},  
\end{equation}
where $a_n\in[0,1]$ determines the relative contributions of the complete and incomplete observations.
We assume $a_n\to a$ as $n\to\infty$.
Motivated by \textcite{amro_permuting_2017}, we use $a_n = 2n_1/(n+n_1)$, which accommodates the two commonly applied methods.
In fact, if all observations are complete ($n_1=n$), there is $a_n=1$ and $K_n(\mathbf{x},t)$ reduces to the squared paired $t$-type statistic
\eqref{eq:paired_pointwise_statistics}. 
If no complete pairs are observed ($n_1=0$), we have $a_n=0$ and $K_n(\mathbf{x},t)$ is the squared Welch-type statistic \eqref{eq:missing_pointwise_statistics}.

Global test statistics for functional hypothesis tests are typically obtained by integrating a point-wise statistic or taking its supremum \parencite[e.g.][]{zhang2011statistical, munko_multiple_2025, munko_general_2025}. 
We follow this idea to obtain test statistics for inferring the hypotheses in Equation~(\ref{eq:hypothesis}) and first consider the supremum of the point-wise weighted statistic 
\begin{align}
		S_n(\mathbf{x}) &=  \sup_{t \in \mathcal{T}}K_n(\mathbf{x},t). \label{eq:T_max}
\end{align}
Moreover, if $\mathcal T=[b,c]\subset\mathbb R,\, b<c,$ is a compact interval, we additionally consider the integral-type statistic
\begin{align}
	I_n(\mathbf{x}) &=   \int_{\mathcal{T}} K_n(\mathbf{x},t) dt. \label{eq:T_int}
\end{align}
Here and throughout, whenever an integral-type statistic is considered, $\mathcal T=[b,c]\subset\mathbb R$ is additionally assumed to be a compact interval.
For the construction of global hypothesis tests, we examine the behaviour of these test statistics under the null $\mathcal{H}_0$ and the alternative $\mathcal{H}_1$ in Equation (\ref{eq:hypothesis}).

\begin{satz}\label{thm:distribution_test_statistics}
	Assume Assumption~\ref{thm:assumptions}.
	\begin{enumerate}
		\item Under $\mathcal{H}_0$, the test statistics $I(\mathbf{x})$ and $S(\mathbf{x})$ converge in distribution:
		\begin{align*}
			I_n(\mathbf{x})&\stackrel{d}{\longrightarrow}Z^I=\int_{\mathcal{T}}\sqrt{a}Z_1^2(t)+\sqrt{1-a}Z_2^2(t)dt,\\
			S_n(\mathbf{x})&\stackrel{d}{\longrightarrow}Z^S=\sup_{t\in\mathcal{T}}\left(\sqrt{a}Z_1^2(t)+\sqrt{1-a}Z_2^2(t)\right),
		\end{align*}
		where $	Z_1\sim \GP(0,\cov(Z_1))$ and $Z_2\sim \GP(0,\cov(Z_2))$ are independent Gaussian processes with $\var(Z_i)=1$, $i\in\{1,2\}$ and 
		\begin{align*}
			\cov(Z_1(s),Z_1(t))
			=\,\frac{\gamma_{1-2}(s,t)}{\sqrt{\gamma_{1-2}(s,s)\gamma_{1-2}(t,t)}},
	    \end{align*}
	      \begin{align*}
	    	\cov(Z_2(s),Z_2(t)) =\,\frac{\kappa_3\gamma_1(s,t)+\kappa_2\gamma_2(s,t)}{\sqrt{\kappa_3\gamma_1(t,t)+\kappa_2\gamma_2(t,t)}\sqrt{\kappa_3\gamma_1(s,s)+\kappa_2\gamma_2(s,s)}}
	    \end{align*}
	  for all $s,t\in\mathcal{T}$.
		\item Under $\mathcal{H}_1$, $I_n(\mathbf{x})$ and $S_n(\mathbf{x})$ converge in probability to $\infty$.
	\end{enumerate}
\end{satz}

\textcite{smaga_note_2020} has proven Theorem~\ref{thm:distribution_test_statistics} for the case $n=n_1$ based on stochastic processes valued in $L^2$. 
In Section~\ref{sec:proof_distribution_test_statistics}, we present a detailed proof based on stochastic processes valued in $C^2(\mathcal{T})$ for our more general case.
Let $z^I_{1-\alpha}$ and $z^S_{1-\alpha}$ denote the $(1-\alpha)$-quantiles of $Z^I$ respective $Z^S$.
Theorem~\ref{thm:distribution_test_statistics} shows that $\varphi_n^I=\mathds{1}\{I_n(\mathbf{x})>z^I_{1-\alpha}\}$ and $\varphi_n^S=\mathds{1}\{S_n(\mathbf{x})>z^S_{1-\alpha}\}$ are of asymptotic level $\alpha$ under $\mathcal{H}_0$ and consistent for the alternative $\mathcal{H}_1$. 
However, both limiting null distributions depend on unknown covariance functions and therefore critical values are not directly computable. We therefore propose to approximate the null distributions by resampling, particularly permutation and parametric bootstrap approaches.

\subsection{Resampling Procedures}\label{sec:resampling}
To calculate critical values, we consider two approaches. 
The first is an adoption of the permutation method proposed by \textcite{amro_permuting_2017} for the classical matched pairs setting with missing values. 
The second approach is an adoption of the parametric bootstrap procedure as discussed in \textcite{smaga_note_2020,munko_multiple_2025} for multivariate functional data respective fully observed paired data.

\subsubsection{Permutation procedure}\label{sec:permutation}
The idea introduced by \textcite{amro_permuting_2017} involves permuting the complete and incomplete cases separately and then combining the results using the weighted test statistic as follows: 
for the complete observations $\mathbf{x}$, the idea is to randomly permute components of every matched pair, i.e., we consider i.i.d. permutations $\tau_j:\,\{1,2\}\to\{1,2\},$ $j\in\{1,\dots,n_1\}$, which are uniformly distributed on the symmetric group $\mathcal{S}_2$. 
The random variables $\tau_j$, $j\in\{1,\dots,n_1\}$, are assumed to be independent of $\mathbf{x}=(\mathbf{x}^{(c)},\mathbf{x}^{(m)})$, leading to a permuted sample of complete pairs:
\begin{align}
	\label{eq:pairedpermdata}
	\mathbf{x}^{(c)}_\tau = \begin{pmatrix}
		x_{\tau_1(1)}^{(c)}(t)\\
		x_{\tau_1(2)}^{(c)}(t)
	\end{pmatrix}
	,\dots, 
	\begin{pmatrix}
		x^{(c)}_{\tau_{n_1}(1)}(t)\\
		x^{(c)}_{\tau_{n_1}(2)}(t)    
	\end{pmatrix}, \quad t \in \mathcal{T}.
\end{align}
Then, the paired permutation test statistic $T_{n_1}(\mathbf{x}_\tau^{(c)}, \boldsymbol{\hat\Gamma}_\tau^{(c)})(t)$, $t\in\mathcal{T}$, is calculated as in Equation~(\ref{eq:paired_pointwise_statistics}) using the empirical mean and covariance functions $\boldsymbol{\hat\eta}^{(c)}_\tau$ and $\hat\gamma^{(c)}_{1-2,\tau}$ of the permuted complete pairs, respectively.
For the incomplete observations $\mathbf{x}^{(m)}$, we pool all incomplete observations
\begin{align*}
	\left(y_{1}(t),\dots,y_{n_2+n_3}(t)\right)= \left(x_{11}^{(m)}(t),\dots,x_{1n_2}^{(m)}(t),x_{21}^{(m)}(t),\dots,x_{2n_3}^{(m)}(t)\right),\  t \in \mathcal{T},
\end{align*}
and permute them uniformly at random. 
Mathematically, this means we use a random permutation $\pi:\{1,\dots,n_2+n_3\}\to\{1,\dots,n_2+n_3\}$ that is uniformly distributed on $\mathcal{S}_{n_2+n_3}$ and independent of all other variables to obtain the permuted sample
\begin{align}
	\label{eq:unpairedpermdata}
	\mathbf{x}^{(m)}_\pi =  \left(y_{\pi(1)}(t),\dots,y_{\pi(n_2+n_3)}(t)\right),\,t\in\mathcal{T}.
\end{align}
Note that the whole functions $y_j(t)$, $t\in\mathcal{T}$, $j\in\{1,\dots,n_2+n_3\}$, are permuted, not just their values at a specific point $t$. 
We then compute the Welch-type statistic \eqref{eq:missing_pointwise_statistics} based upon the permuted sample denote by 
$T_{n_2+n_3}(\mathbf{x}_\pi^{(m)}, \boldsymbol{\hat{\Gamma}}_{0,\pi}^{(m)})(t)$ for every $t\in\mathcal{T}$.
Finally, we obtain a permutation version of the weighted test statistic $K_n(\mathbf{x},t)$ in Equation~(\ref{eq:pointwise_weighted}) by combining both in
\begin{align*} 
	K^*_n(\mathbf{x},t)=\sqrt{a_n}\,T^2_{n_1}\left(\mathbf{x}_\tau^{(c)}, \boldsymbol{\hat\Gamma}^{(c)}_\tau\right)(t) + \sqrt{1-a_n} T^2_{n_2+n_3}\left(\mathbf{x}_\pi^{(m)}, \boldsymbol{\hat\Gamma}_{0,\pi}^{(m)}\right)(t),\  t\in\mathcal{T}.
\end{align*}
From this, we derive the permuted versions of the integral- and supremum-type statistics 
\begin{align}
	I^*_n(\mathbf{x}) &=   \int_{\mathcal{T}} K^*_n(\mathbf{x},t) dt, \label{eq:perm_T_int} \\
	S^*_n(\mathbf{x}) &=  \sup_{t \in \mathcal{T}}K^*_n(\mathbf{x},t). \label{eq:perm_T_max}
\end{align}
The following Theorem states the asymptotic behaviour of both.
\begin{satz}
	\label{thm:distribution_perm_test_statistic}
	Suppose Assumption~\ref{thm:assumptions}.
	Then, the resampling test statistics $I^*_n(\mathbf{x})$ and $S^*_n(\mathbf{x})$ given in Equations~(\ref{eq:perm_T_int}) and (\ref{eq:perm_T_max}) always converge, conditionally given the data in probability, in distribution to some random variable, i.e., we have under $\mathcal{H}_0$ as well as under $\mathcal{H}_1$ in (\ref{eq:hypothesis}):
	\begin{align*}
		\sup_{w\in\br}&\left|\prob\left(I_n^*(\mathbf{x})\le w|\mathbf{x}\right)-\prob\left(\int_\mathcal{T}\sqrt{a}Z_1^2(t)+\sqrt{1-a}Z_{2,\pi}^2(t)dt\le w\right)\right|\stackrel{p}{\longrightarrow}0,\\
		\sup_{w\in\br}&\left|\prob\left(S_n^*(\mathbf{x})\le w|\mathbf{x}\right)-\prob\left(\sup_{t\in\mathcal{T}}\left(\sqrt{a}Z_1^2(t)+\sqrt{1-a}Z_{2,\pi}^2(t)\right)\le w\right)\right|\stackrel{p}{\longrightarrow}0,
	\end{align*}
	where the concrete form of the Gaussian process $Z_{2,\pi}$ is given in Section~\ref{sec:proof_distribution_perm_test_statistics}.
\end{satz}
Theorem~\ref{thm:distribution_perm_test_statistic} completely characterizes the asymptotic conditional distribution of the permutation statistics $I^*_n(\mathbf{x})$ and $S^*_n(\mathbf{x})$ given in Equations~(\ref{eq:perm_T_int}) and (\ref{eq:perm_T_max}). 
Comparing with Theorem~\ref{thm:distribution_test_statistics}, it is apparent that the limiting distributions do not coincide with the null distribution of the respective test statistics in general. 
The reason for this comes from the Welch-type component, whose permutation limit differs from its null limit. 
Consequently, the proposed permutation procedure is in general not asymptotically exact without additional assumptions. 
The two following corollaries give conditions for asymptotic correctness.

\begin{cor}\label{rem:special_cases}
	If either $\kappa_2=\kappa_3$ (balancedness) or $\gamma_1=\gamma_2$ (homoscedasticity), the distributions of $Z_{2,\pi}$ and $Z_2$ coincide and the permutation procedures based upon $I^*_n(\mathbf{x})$ and $S^*_n(\mathbf{x})$ are asymptotically valid, i.e.
	\begin{align}\label{eq:perm:consistency I}
		\sup_{w\in\br}&\left|\prob\left(I_n^*(\mathbf{x})\le w|\mathbf{x}\right)-\prob\left(\int_\mathcal{T}\sqrt{a}Z_1^2(t)+\sqrt{1-a}Z_{2}^2(t)dt\le w\right)\right|\stackrel{p}{\longrightarrow}0,\\\label{eq:perm:consistency S}
		\sup_{w\in\br}&\left|\prob\left(S_n^*(\mathbf{x})\le w|\mathbf{x}\right)-\prob\left(\sup_{t\in\mathcal{T}}\left(\sqrt{a}Z_1^2(t)+\sqrt{1-a}Z_{2}^2(t)\right)\le w\right)\right|\stackrel{p}{\longrightarrow}0.
	\end{align}
\end{cor}
The next corollary can be seen as a special case of the former. 
It is stated separately as it provides the first theoretical justification of the permutation procedure proposed by \textcite{smaga_note_2020}.

\begin{cor}\label{rem:complete_case}
	If $n=n_1$ (no missing data), Equations ~(\ref{eq:perm:consistency I}) and~(\ref{eq:perm:consistency S}) hold. 
	Consequently, the permutation procedure is asymptotically valid in the complete-data setting.
\end{cor}

The two corollaries identify settings in which the conditional permutation distribution consistently approximates the null distribution
of the corresponding test statistic. 
This is the case if  $\kappa_2=\kappa_3$, $\gamma_1=\gamma_2$ or $n=n_1$. 
Under these conditions, the conditional permutation quantiles can be used as asymptotically valid critical values. 
Denote by $q^{*,I}_{1-\alpha}(\mathbf{x})$ and $q^{*,S}_{1-\alpha}(\mathbf{x})$, the empirical $(1-\alpha)$-quantiles of the conditional distribution functions $w\mapsto \prob(I_n^*(\mathbf{x})\le w|\mathbf{x})$ and $w\mapsto \prob(S_n^*(\mathbf{x})\le w|\mathbf{x})$, 
respectively.  
The resulting permutation tests are defined by 
\begin{align}\label{eq:tests}
	\varphi_n^{*,I}=\mathds{1}\left\{I_n(\mathbf{x})>q^{*,I}_{1-\alpha}(\mathbf{x})\right\};\quad \varphi_n^{*,S}=\mathds{1}\left\{S_n(\mathbf{x})>q^{*,S}_{1-\alpha}(\mathbf{x})\right\}.
\end{align}
Under $\mathcal{H}_0$ and any of the above conditions, the conditional critical values $q^{*,I}_{1-\alpha}(\mathbf{x})$ and $q^{*,S}_{1-\alpha}(\mathbf{x})$ converge in probability to the corresponding $(1-\alpha)$-quantiles of the respective test statistics limiting null distributions. Hence, the tests $\varphi_n^{*,I}$ and $\varphi_n^{*,S}$, are asymptotically exact by 
Lemma~1 in \textcite{janssen_how_2003}.
Moreover, under $\mathcal{H}_1$, the conditional critical values converge in probability to some finite constants 
by Theorem~\ref{thm:distribution_perm_test_statistic}, whereas the test statistics diverge to infinity by Theorem~\ref{thm:distribution_test_statistics}. 
Therefore, the tests $\varphi_n^{*,\ell}$, $\ell\in\{I,S\}$, are consistent due to Theorem~7 in \textcite{janssen_how_2003}.

In practice, the conditional critical values are approximated by Monte Carlo:
for $B$ permutation samples the estimated $p$-value of the proposed permutation tests $\varphi_n^{*,\ell}$, $\ell\in\{I,S\}$, can be computed by
\begin{align}\label{eq:perm_pvalues}
	p^{*,\ell} =\frac{1}{B+1}\left(1+\sum_{b=1}^{B} \mathds{1}\left\{\ell_{n,b}^*(\mathbf{x})\geq \ell_n(\mathbf{x})\right\}\right)
\end{align}
following Equation~(17.7) in \textcite{lehmann_testing_2022}.
A concrete algorithm for computing Equation~(\ref{eq:perm_pvalues}) is explained in \texttt{additional.pdf} of the Supplementary Material \parencite{baumeister_supplementary_2026}.

In the general case, the conditional permutation distributions do not approximate the null distribution of $I_n$ or $S_n$. 
This is in contrast to the point-wise application considered in \textcite{amro_permuting_2017}. 
As the functional permutation tests are therefore not asymptotically valid without additional assumptions, we consider another resampling technique that will turn out to be asymptotically valid without additional assumptions.

\subsubsection{Parametric bootstrap procedure}\label{sec:parametric}

The parametric bootstrap is an asymptotic model-based bootstrap procedure that has been successfully applied to non-functional data in matched pairs and general \textsc{manova} and factorial designs, e.g., \textcite{konietschke_parametric_2015, friedrich_mats_2018, zimmermann_multivariate_2020, baumeister_multivariate_2026, amro_asymptotic-based_2021}. 
It has also been used for functional data for various setups \parencite[e.g.][]{cuevas2004anova, martinez2011repeated, martinez2012general, smaga_note_2020, munko_general_2025, munko_multiple_2025}.
The idea is to generate the bootstrap sample from centred Gaussian process whose covariance function  $\boldsymbol{\hat{\Gamma}}$ is estimated from the observed data.
For the complete observations $\mathbf{x}^{(c)}$, we generate
\begin{align}
	\label{eq:bootstrap_data_paired}
	\mathbf{x}_{j,\mathcal{P}}^{(c)}\sim \GP_2 \left(\mathbf{0},\boldsymbol{\hat{\Gamma}}^{(c)}\right),\,j\in\{1,\dots,n_1\}.
\end{align}
Note that this preserve the paired structure and mimics the covariance structure of the functional data.
In contrast, no pairing information is available for the incomplete observations. 
Hence, we generate the incomplete samples independently according to
\begin{align}
	\label{eq:bootstrap_data_unpaired}
	x_{j,\mathcal{P}}^{(m)}\sim
	\begin{cases}
		\GP\left(0,\hat{\gamma}_{1}^{(m)}\right),\,j\in\{1,\dots,n_2\},\\
		\GP\left(0,\hat{\gamma}_{2}^{(m)}\right),\,j\in\{1,\dots,n_3\},
	\end{cases}
\end{align}
where the covariance estimators $\hat{\gamma}_{1}^{(m)}$ and $\hat{\gamma}_{2}^{(m)}$ are calculated as in Equation~(\ref{eq:empcovariance}). 
Similar to the permutation approach, the bootstrap point-wise statistic is then calculated as 
\begin{align*}
	K^\mathcal{P}_n(\mathbf{x},t)=\sqrt{a_n}\,T^2_{n_1}\left(\mathbf{x}_\mathcal{P}^{(c)}, \boldsymbol{\hat\Gamma}_\mathcal{P}^{(c)}\right)(t) + \sqrt{1-a_n} T^2_{n_2+n_3}\left(\mathbf{x}_\mathcal{P}^{(m)}, \boldsymbol{\hat\Gamma}_{0,\mathcal{P}}^{(m)}\right)(t),\  t\in\mathcal{T}.
\end{align*}
Here, $\boldsymbol{\hat\Gamma}_\mathcal{P}^{(c)}$ and $\boldsymbol{\hat\Gamma}_{0,\mathcal{P}}^{(m)}$ are the parametric bootstrap version of $\boldsymbol{\hat\Gamma}^{(c)}$ and $\boldsymbol{\hat\Gamma}_{0}^{(m)}$, respectively.
The corresponding global bootstrap statistics are
\begin{align}
	I^\mathcal{P}_n(\mathbf{x}) &=   \int_{\mathcal{T}} K^\mathcal{P}_n(\mathbf{x},t) dt, \label{eq:boot_T_int} \\
	S^\mathcal{P}_n(\mathbf{x}) &=  \sup_{t \in \mathcal{T}}K^\mathcal{P}_n(\mathbf{x},t). \label{eq:boot_T_max}
\end{align}
The following theorem establishes their asymptotic behaviour. 
\begin{satz}
	\label{thm:distribution_boot_test_statistic}
		Suppose Assumption~\ref{thm:assumptions}.
		Then, the resampling test statistics $I^\mathcal{P}_n(\mathbf{x})$ and $S^\mathcal{P}_n(\mathbf{x})$ given in Equations~(\ref{eq:boot_T_int}) and (\ref{eq:boot_T_max}) always converge, conditionally given the data in probability, in distribution to the same random variable, to which $I_n(\mathbf{x})$ or rather $S_n(\mathbf{x})$ converge under $\mathcal{H}_0$ (\ref{eq:hypothesis}), i.e., we have under $\mathcal{H}_0$ as well as under $\mathcal{H}_1$:
	\begin{align*}
		\sup_{w\in\br}&\left|\prob\left(I_n^\mathcal{P}(\mathbf{x})\le w|\mathbf{x}\right)-\prob\left(\int_\mathcal{T}\sqrt{a}Z_1^2(t)+\sqrt{1-a}Z_2^2(t)dt\le w\right)\right|\stackrel{p}{\longrightarrow}0,\\
		\sup_{w\in\br}&\left|\prob\left(S_n^\mathcal{P}(\mathbf{x})\le w|\mathbf{x}\right)-\prob\left(\sup_{t\in\mathcal{T}}\left(\sqrt{a}Z_1^2(t)+\sqrt{1-a}Z_2^2(t)\right)\le w\right)\right|\stackrel{p}{\longrightarrow}0.
	\end{align*}
\end{satz}
In contrast to the permutation procedure, the bootstrap approximation is asymptotically valid without additional assumptions such as balanced missingness or identical covariance operators. 
Therefore, with the same arguments as in Subsection~\ref{sec:permutation}, the bootstrap tests
\begin{align}\label{eq:boot_tests}
	\varphi_n^{\mathcal{P},I}=\mathds{1}\left\{I_n(\mathbf{x})>q^{\mathcal{P},I}_{1-\alpha}(\mathbf{x})\right\};\quad \varphi_n^{\mathcal{P},S}=\mathds{1}\left\{S_n(\mathbf{x})>q^{\mathcal{P},S}_{1-\alpha}(\mathbf{x})\right\},
\end{align}
are asymptotically exact under the null hypothesis and consistent against fixed alternatives. 
Here, $q^{\mathcal{P},I}_{1-\alpha}(\mathbf{x})$ and $q^{\mathcal{P},S}_{1-\alpha}(\mathbf{x})$ denote the empirical $(1-\alpha)$-quantiles of the conditional distribution functions $w\mapsto \prob(I_n^\mathcal{P}(\mathbf{x})\le w|\mathbf{x})$ and $w\mapsto \prob(S_n^\mathcal{P}(\mathbf{x})\le w|\mathbf{x})$, respectively.

Again, $p$-values can be computed via Monte-Carlo: for $B$ bootstrap samples, the estimated $p$-value of the bootstrap tests $\varphi_n^{\mathcal{P},\ell}$, $\ell\in\{I,S\}$, are given by
\begin{align}\label{eq:boot_pvalues}
	p^{\mathcal{P},\ell} = \frac{1}{B+1}\left(1+\sum_{b=1}^{B} \mathds{1}\left\{\ell_{n,b}^\mathcal{P}(\mathbf{x})>\ell_n(\mathbf{x})\right\}\right).
\end{align}
An algorithm for computing the $p$-values in Equation~(\ref{eq:boot_pvalues})  is explained in \texttt{additional.pdf} of the Supplementary Material \parencite{baumeister_supplementary_2026}.

\section{Simulation Studies}\label{sec:simulation}
In this section, we investigate the finite sample behaviour of the proposed permutation and parametric bootstrap tests by means of an extensive simulation study. 
It includes an evaluation of the type I error control under the null hypothesis (at a chosen level of $\alpha = 0.05$) as well as an analysis of power under different alternatives. 
We compare our proposed methods with complete-case and imputation-based approaches. 
For the complete-case procedures, we use the permutation methods $I_{Pc}$ and $S_{Pc}$ of \textcite{smaga_note_2020} based on an integral- or supremum-type statistic, respectively, as well as the bootstrap approach $CC_c$ of \textcite{crainiceanu2012bootstrap}, all applied to the complete data set only. 
For the imputation-based approaches, called $I_{Pi}$, $S_{Pi}$, and $CC_i$, these methods are combined with the imputation method for functional data proposed by \textcite{jang_bayesian_2021}.
The abbreviations $I_{P}$, $S_{P}$, $I_{PB}$, and $S_{PB}$ indicate the newly proposed methods based on permutation ($P$) or parametric bootstrapping ($PB$) combined with the integral- ($I$) or supremum-type ($S$) statistic. 
The competing methods are summarized in Table~\ref{tab:compared_methods}.
\begin{table}[t]
	\caption{List of methods that are compared in the simulation study.}
	\begin{tabularx}{\textwidth}{lXX}
		\toprule
		Abbr. & Citation & Description \\
		\midrule
		$I_{Pc}$, $S_{Pc}$ & \textcite{smaga_note_2020} & permutation tests applied to complete data pairs\\
		$CC_c$ & \textcite{crainiceanu2012bootstrap} & bootstrap test applied to complete data pairs\\
		\midrule[0.01pt]
		$I_{Pi}$, $S_{Pi}$ & \textcite{smaga_note_2020,jang_bayesian_2021} & permutation tests applied to imputed data\\
		$CC_i$ & \textcite{crainiceanu2012bootstrap, jang_bayesian_2021} & bootstrap test applied to imputed data\\
		\midrule[0.01pt]
		$I_{P}$,  $S_{P}$ & -- & new permutation tests applied on the whole data\\
		$I_{PB}$, $S_{PB}$ & -- & new parametric bootstrap tests based applied on the whole data\\
		\bottomrule
	\end{tabularx}
	\label{tab:compared_methods}
\end{table}

\subsection{Simulation Setup}
Our simulation setup is inspired by the simulation study conducted by  \textcite{smaga_note_2020} for the case of paired functional data without missing values. 
We simulate paired functional samples using the following framework:
\begin{align*}
	x_{ij}(t)&= c_i(t) + e_{ij}(t)
\end{align*}
for $t \in [0,1]$, $i \in \{1,2\}$ and $j\in\{1,\dots,n\}$.

\paragraph{Mean functions.}
We consider the following seven different models, each representing specific forms of mean functions $c_1$ and $c_2$:
\begin{align*}
	\text{Model } M1: &\  c_1(t) = c_2(t) = (\sin(2\pi t^2))^5 I_{[0,1]}(t), \\
	\text{Model } M2: &\  c_1(t) = (\sin(2\pi t^2))^5 I_{[0,1]}(t), \  c_2(t) = (\sin(2\pi t^2))^7 I_{[0,1]}(t), \\
	\text{Model } M3: &\  c_1(t) = (\sin(2\pi t^2))^5 I_{[0,1]}(t), \  c_2(t) = (\sin(2\pi t^2))^3 I_{[0,1]}(t),\\
	\text{Model } M4: &\  c_1(t) = c_2(t) = \sqrt{6t/\pi}e^{-6t} I_{[0,1]}(t), \\
	\text{Model } M5: &\  c_1(t) = \sqrt{6t/\pi}e^{-6t} I_{[0,1]}(t), \  c_2(t) = \sqrt{13t/(2\pi)}e^{-13t/2} I_{[0,1]}(t), \\
	\text{Model } M6: &\  c_1(t) =  \sqrt{6t/\pi}e^{-6t} I_{[0,1]}(t), \  c_2(t) = \sqrt{11t/(2\pi)}e^{-11t/2} I_{[0,1]}(t), \\
	\text{Model } M7: &\  c_1(t) = \sqrt{6t/\pi}e^{-6t} I_{[0,1]}(t), \  c_2(t) = \sqrt{5}t^{2/3}e^{-7t} I_{[0,1]}(t).
\end{align*}
Consequently, the null hypothesis holds under Models $M1$ and $M4$, while the alternative hypothesis holds under Models $M2$, $M3$, and $M5 - M7$. 
The first three models have first been studied by \textcite{smaga_note_2020}, while the remaining models were also investigated by \textcite{martinez2011repeated}.

\paragraph{Error processes.}
We consider two different random errors that are defined as follows:
\begin{align}
	\text{Normal:} \  &e_{1j}(t) = wB_{1j}(t), \  e_{2j}(t) = \rho e_{i1}(t) + w\sqrt{1-\rho^2}B_{2j}(t), \label{eq:simu_data_normal_errors}\\
	\text{Lognormal:} \  &e_{1j}(t) = \exp\left(wB_{1j}(t)\right), \  e_{2j}(t) = \exp\left(\rho e_{1j}(t) + w\sqrt{1-\rho^2}B_{2j}(t)\right),\label{eq:simu_data_lognormal_errors}
\end{align}
where $t \in [0,1]$, $j\in\{1,\dots,n\}$. 
Thereby, three cases of $\rho\in\{0.25,0.5,0.75\}$ corresponding to low, moderate, and high correlations are considered.
For the Models $M1-M3$, the error terms (\ref{eq:simu_data_normal_errors}) and (\ref{eq:simu_data_lognormal_errors}) are considered with $w=0.5$. 
For Models $M4-M7$, we use the same error functions from the simulation study in \textcite{martinez2011repeated}, adjusting our error terms to be ten times smaller by using $w=1/20$ in the respective formulas in Equations~(\ref{eq:simu_data_normal_errors}) and~(\ref{eq:simu_data_lognormal_errors}). 
For $B_{ij}(t)$, $i\in\{1,2\}$, $j\in\{1,\dots,n\}$, we explore different choices. 
First, equivalently to \textcite{martinez2011repeated}, $B_{i1}$ and $B_{i2}$ are two independent standard Brownian Bridges. 
We call this the Brownian bridges' (BB) case. 
Second, we consider co-called Zhang processes, which have been introduced in \textcite{zhang2011statistical} and are defined as follows for every $i\in\{1,2\}$ and $j\in\{1,\dots,n\}$:
\begin{alignat*}{2}
	B_{ij}(t)&=\mathbf{b}_{ij}^{\top}\boldsymbol{\Psi}(t), & \\
	\mathbf{b}_{ij}&=(b_{ij1},\dots,b_{ij7})^{\top}, & \quad \boldsymbol{\Psi}(t)&=(\Psi_1(t),\dots,\Psi_7(t))^{\top},\\
	b_{ijr}&=\sqrt{\lambda_{r}}v_{ijr},\,r\in\{1,\dots,7\},&\quad   \Psi_r(t)&=\begin{cases}
		1,\; r=1,\\
		\sqrt{2}\sin(2\pi rt),\; r\in\{2,4,6\},\\
		\sqrt{2}\cos(2\pi rt),\;r\in\{3,5,7\},
	\end{cases}\\
	\lambda_{r}&=0.3\rho_z^r,\,v_{ijr}\stackrel{i.i.d.}{\sim} \sqrt{\frac{3}{5}}t_5, & \\
	\rho_z&\in\{0.1,0.5,0.9\}. & 
\end{alignat*}
Here, $t_5$ is the $t$-distribution with $5$ degrees of freedom. 
The parameter $\rho_z$ corresponds to the correlation of the functional data. 
In this framework, the correlation between measurements at different time points within a single functional observation differs. 
The smaller the value of $\rho_z$, the higher the correlation.

Note, that the asymptotic validity condition of the permutation procedures given in Corollary~\ref{rem:special_cases} holds for all simulation settings with Brownian Bridges and Zhang processes.
Therefore, we consider some heteroscedastic settings that are also based on Brownian Bridges.
We set $B_{1j}=Z_{1j}$ and $B_{2j}=\sigma Z_{2j}$, where $\sigma\in\{\sqrt{2},\sqrt{3}\}$ and $Z_{ij}$ are standard Brownian Bridges for $i\in\{1,2\}$ and $j\in\{1,\dots,n\}$.
With these processes, we consider the models $M1$, $M2$, $M4$, $M5$, the normal errors in Equation~(\ref{eq:simu_data_normal_errors}) and all correlations $\rho\in\{0.25,0.5,0.75\}$.
On these heteroscedastic settings, we consider only the methods $I_{Pc}$, $S_{Pc}$, $I_P$, $S_P$, $I_{PB}$, and $S_{PB}$.

\paragraph{Sample Sizes and Missingness.}
Missing values has been introduced using the \textsc{mcar} mechanism by randomly inserting them into the first or second component of the functional data, until a fixed number of missing values $n_2$ for the second component and $n_3$ for the first is achieved.
We thereby consider two sample size scenarios. 
In Scenario 1, we consider various balanced and unbalanced sample sizes:
$$(n_1,n_2,n_3) \in \{(5,5,5),(5,5,10),(5,10,10),(10,10,5),(10,5,5),(10,10,10),(20,20,20)\}.$$ 
In Scenario 2, we fix the number of pooled observations $n=n_1+n_2+n_3=30$ and investigate various unbalanced sample size settings with missingness ranging between $30\%$ and $80\%$: 
$$(n_1,n_2,n_3) \in \{(21,5,4), (18,6,6), (15,8,7), (12,9,9), (9,11,10), (6,12,12)\}.$$
In both scenarios, the balancedness condition $n_2=n_3$ from Corollary~\ref{rem:special_cases} holds for some of the considered sample sizes.
For the heteroscedastic settings, we consider only the sample sizes of Scenario 2.

\paragraph{Discretization.}
To simulate functional data that reflect real-world scenarios, where observations are typically not continuous, we generate trajectories of the functions $\mathbf{x}_{j}$, $j\in\{1,\dots,n\}$, discretized at design time points $t_1,\dots,t_\iota$. 
The time points are equispaced in the interval $[0,1]$, and we consider $\iota=50$ for BB processes and $\iota\in\{25,50\}$ for Zhang processes.

All simulations were performed using the statistical computing environment \textsf{R}, version 4.5.2~\parencite{r_core_team_r_2025}, with each simulation setting based on $\texttt{nsim} = 1,000$ Monte Carlo runs and $B = 1,000$ permutation or bootstrap runs. 
For the imputation method of \textcite{jang_bayesian_2021}, the Gibbs sampler was run for $10,000$ iterations, with the first $5,000$ samples discarded as burn-in.
Similar parameters are used in the data example of \textcite{jang_bayesian_2021}.
Note that we only used a single imputation based on the procedure of \textcite{jang_bayesian_2021} as its computational burden is very high (approximately two days per simulation setting using a supercomputer with many parallel processes), and because Rubin-type combining rules are not directly applicable to our asymptotically non-pivotal test statistics.

\subsection{Simulation results}
\begin{figure}[!t]
	\centering
	\includegraphics[width=0.99\textwidth]{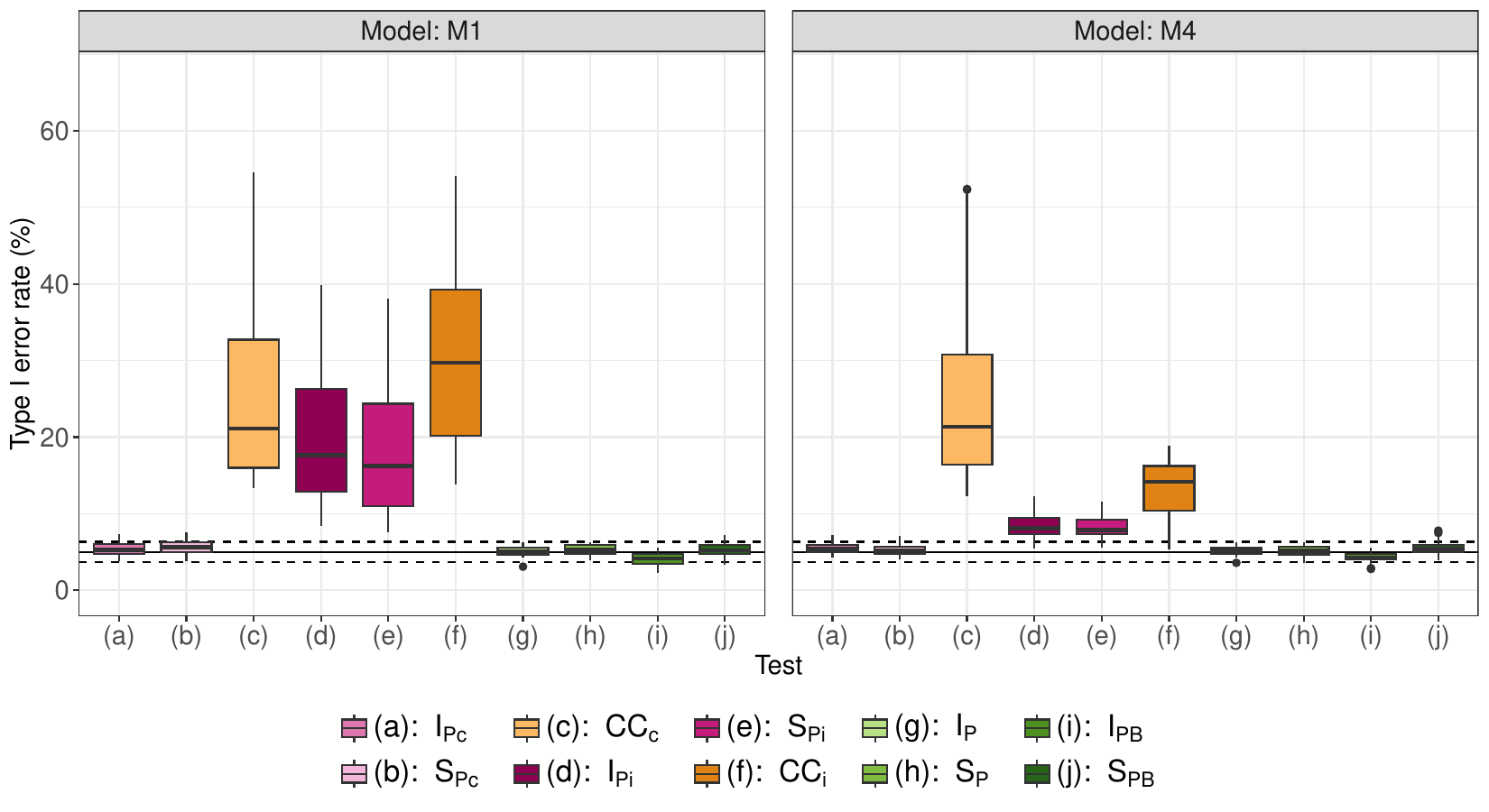}
	\caption{Boxplots for the empirical type I error rate (as percentages) of all tests obtained in Models $M1$ and $M4$ and Scenario 2 with BB processes. The solid line denotes the significance level $\alpha=5\%$, and the dashed line denotes the limits of $95\%$ binomial confidence intervals $[3.6,6.4]$ for 1000 simulation runs \parencite[cf.][]{DuchesneFrancq2015}.}
	\label{fig:BB_m1m4}
\end{figure}

\paragraph{Type I Error Control}
Complete simulation results are provided as raw data in the Supplementary Material \parencite[][\texttt{simuresults\_full.csv}]{baumeister_supplementary_2026}.
In the following, we use boxplots from the \textsf{R}-package \texttt{ggplot2} \parencite{wickham_ggplot2_2016} to visualize the simulation results.
Additionally, further plots are presented in Section~2 of the \texttt{additional.pdf} in the Supplementary Material.
Overall, the methods $CC_c$, $I_{Pi}$, $S_{Pi}$, and $CC_i$ exhibit a too liberal behaviour, while the other methods have empirical type I error rates close to the nominal level of $5\%$.
This pattern is illustrated in Figure~\ref{fig:BB_m1m4} for the BB processes under Scenario 2 and models $M1$ and $M4$. 
For the other simulation settings, similar plots are given in the Supplementary Material.
The liberal behaviour of $CC_c$ for small sample sizes is consistent with the findings of \textcite{crainiceanu2012bootstrap}, however, this method can perform well for larger sample sizes.
The liberal behaviour of the imputed methods $I_{Pi}$, $S_{Pi}$, and $CC_i$ could also be related to the comparatively high proportions of incomplete observations in our simulation settings.
In particular, the larger the missingness, the harder it is to impute them. 
\textcite{jang_bayesian_2021} considered missing rates up to $30\%$, which corresponds to the smallest proportion considered in Scenario~2. 
This can be an explanation for the relatively good type I error control of $I_{Pc}$ and $S_{Pc}$, while the corresponding imputation versions $I_{Pi}$ and $S_{Pi}$ are rather liberal.
Only in the special case of Model $M4$, the methods $I_{Pi}$, $S_{Pi}$, and $CC_i$ are less liberal even control the type I error rate reasonably well for some cases. Nevertheless, for most simulations settings, the type I error control of the imputation-based procedures seems to be unacceptably liberal.
Consequently, procedures based either on the complete cases with $CC_c$ or on imputed functional observations with $I_{Pi}$, $S_{Pi}$, and $CC_i$ should be used with caution in small samples or when the proportion of incomplete observations is large.
\begin{figure}[!ht]
	\centering
	\includegraphics[width=\textwidth]{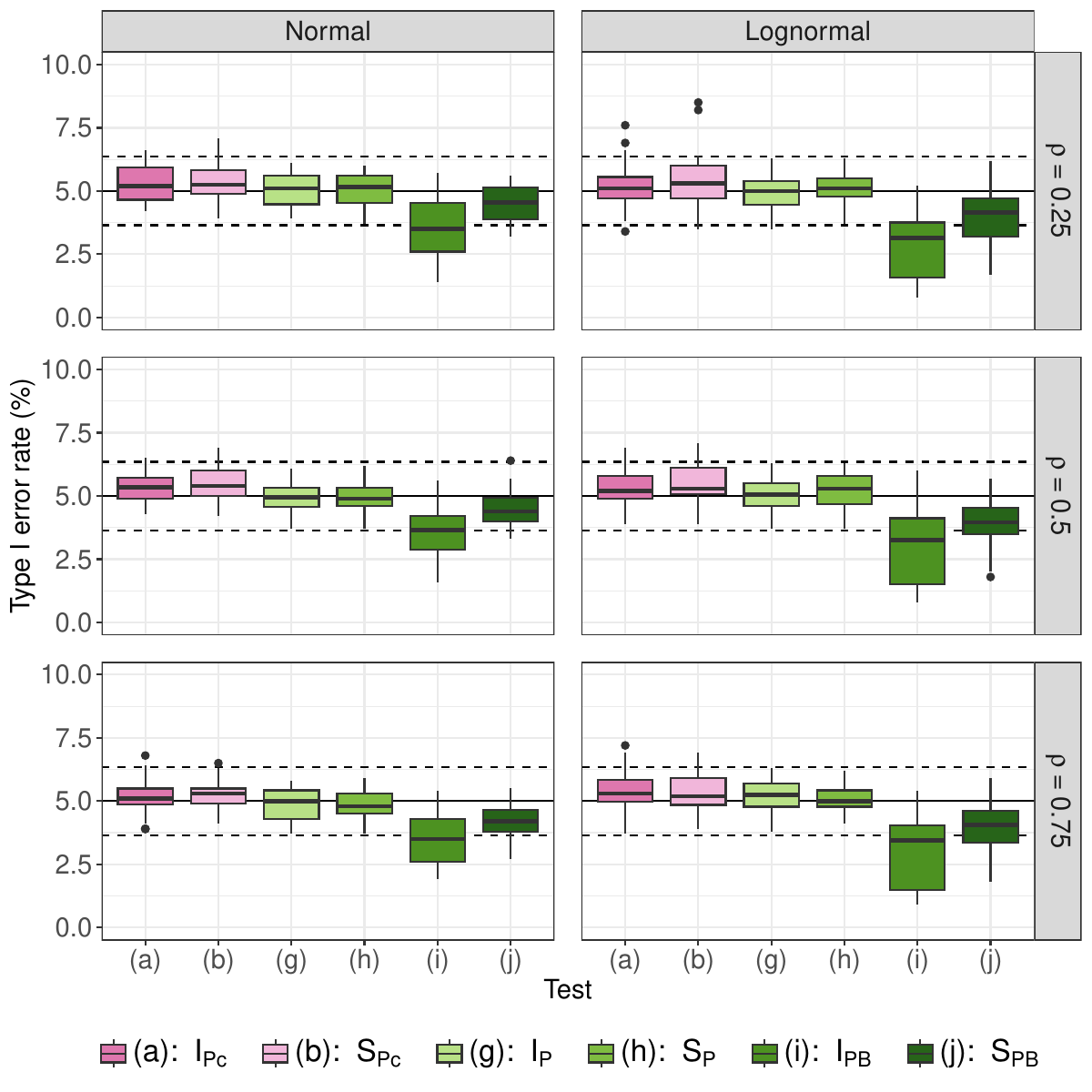}
	\caption{Boxplots for the empirical type I error rate (as percentages) of selected tests obtained in Model $M1$ with Scenario 2 and Zhang processes differentiated by the considered correlations and distributions. The solid line denotes the significance level $\alpha=5\%$, and the dashed line denotes the limits of $95\%$ binomial confidence intervals $[3.6,6.4]$ for 1000 simulation runs \parencite[cf.][]{DuchesneFrancq2015}.}
	\label{fig:Zhang_m1_scenario2_grid}
\end{figure}
However, the complete-pair permutation procedures $I_{Pc}$ and $S_{Pc}$ control the type I error rate in most cases and are liberal for some settings with very few complete pairs, particularly when ($n_1\in\{5,6\}$), and high correlations ($\rho=0.75$).
In contrast, the new methods have a satisfactory type I error control. 
Only the parametric bootstrap integral test $I_{PB}$ tends to be slightly conservative for some scenarios, whereas $S_{PB}$ sometimes exhibits a slightly liberal behaviour, see, e.g. Figure~\ref{fig:BB_m1m4} as well as Section~2 of the \texttt{additional.pdf} in Supplementary Material \parencite{baumeister_supplementary_2026}.

Figure~\ref{fig:Zhang_m1_scenario2_grid} presents a grid of boxplots of the empirical type I error rates under Model \(M1\), Scenario~2, and the Zhang processes, stratified by error distributions and the correlations. 
For a better overview, the liberal methods $CC_c$, $I_{Pi}$, $S_{Pi}$, and $CC_i$ are omitted. 
The conservative behaviour of $I_{PB}$ is visible overall and is more pronounced in the lognormal settings.  
$S_{PB}$ tends to be slightly liberal, though most results remain in the binomial interval. 
In other settings, e.g., with model $M1$, Scenario 1, and Zhang processes, $S_{PB}$ is slightly conservative but still acceptable. 
Analogous plots for other combination of process and model type as well as sample-size scenario are reported in Supplementary Material.
In none of these settings, $I_{PB}$ is as conservative as in Figure~\ref{fig:Zhang_m1_scenario2_grid}.
\begin{figure}[!t]
	\centering
	\includegraphics[width=\textwidth]{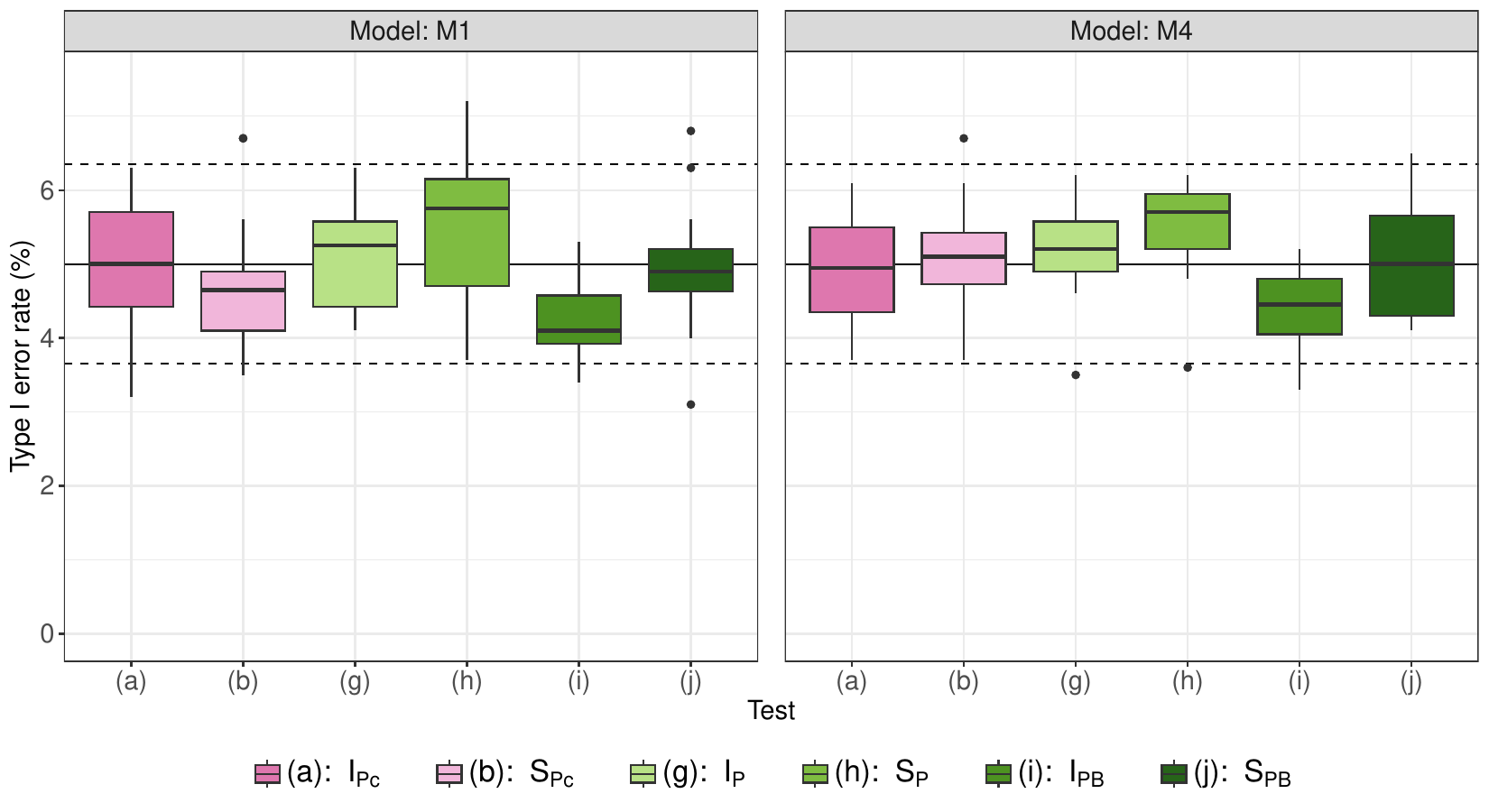}
	\caption{Boxplots for the empirical type I error rate (as percentages) of selected tests obtained in Models $M1$ and $M4$ and the heteroscedastic BB processes on the sample sizes in Scenario 2, where $n_2\not=n_3$ holds. The solid line denotes the significance level $\alpha=5\%$, and the dashed line denotes the limits of $95\%$ binomial confidence intervals $[3.6,6.4]$ for 1000 simulation runs \parencite[cf.][]{DuchesneFrancq2015}.}
	\label{fig:BB_het_paired_unequal samples}
\end{figure}

Note that the permutation procedures in the tests $I_P$ and $S_P$ is valid for BB and Zhang processes, but not for the heteroscedastic BB processes.
To examine the behaviour of the permutation methods $I_P$ and $S_P$ on this data in more detail, Figure~\ref{fig:BB_het_paired_unequal samples} displays results for heteroscedastic BB processes with unbalanced missingness $n_2\not=n_3$.
In these settings the conditions of Corollary~\ref{rem:special_cases} are not satisfied, and Theorem~\ref{thm:distribution_perm_test_statistic} generally yields different limiting distributions of the test statistic and its permutation version. 
However, the empirical type I error rates of $I_P$ and $S_P$ are almost always within the $95\%$ binomial interval. 
This is similar to the behaviour in all other considered simulations settings. 
Specially, in Figure~\ref{fig:BB_het_paired_unequal samples} the behaviour of $I_P$ and $S_P$ is comparable to the asymptotically correct methods $I_{PB}$ and $S_{PB}$.
A possible explanation is that point-wise validity of the permutation approaches is important to provide a good finite-sample approximation for the integral- and supremum-type statistics in the considered settings.
These findings show that the permutation procedures can perform well, even in settings where they are not asymptotically valid. 
They should nevertheless be used with cautiously as simulations cannot replace a theoretical justification.

\paragraph{Power}
\begin{figure}[!t]
	\centering
	\includegraphics[width=\textwidth]{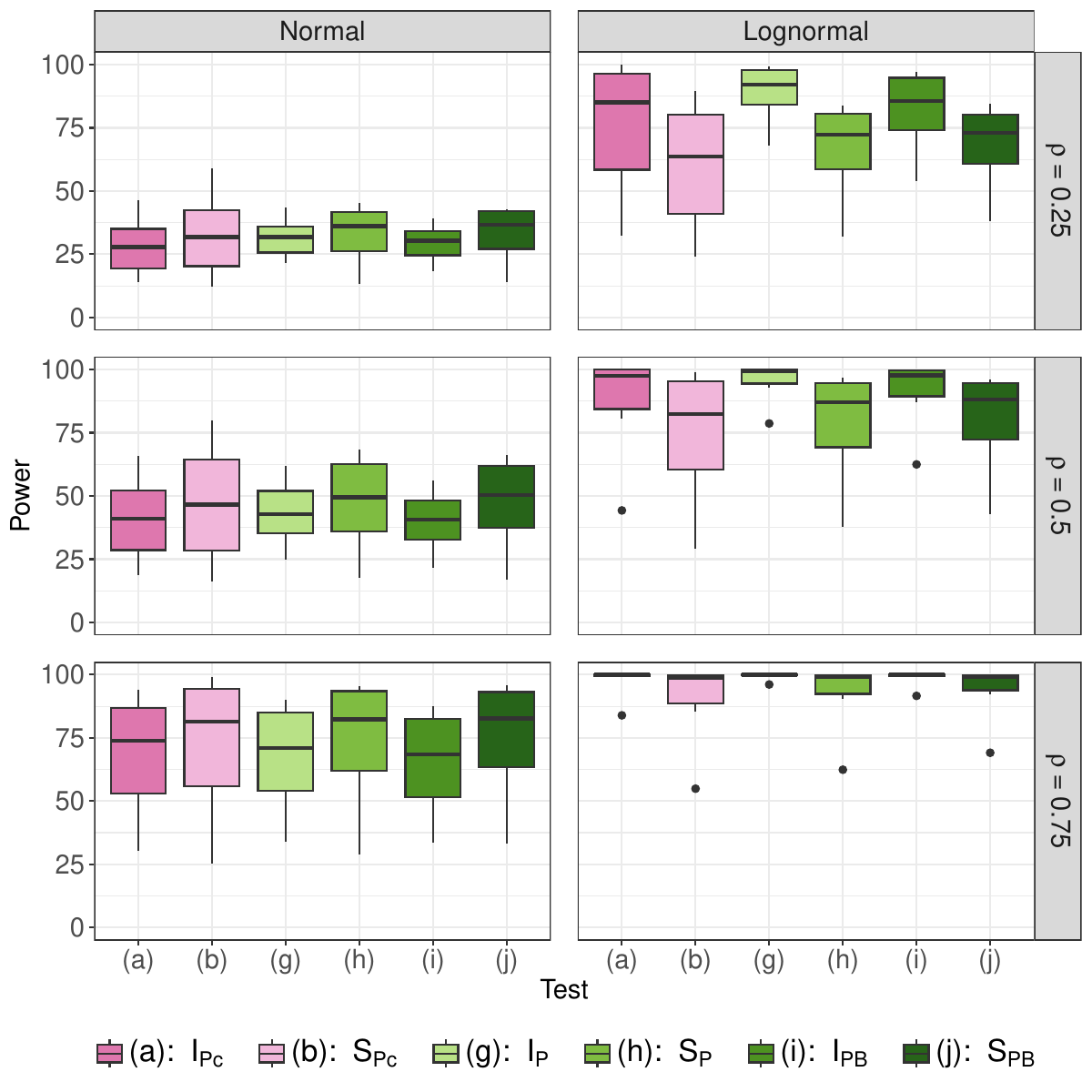}
	\caption{Boxplots for the empirical power (as percentages) of selected tests obtained for Model $M5$ with BB processes in Scenario 2.}
	\label{fig:BB_m5_scenario2_grid}
\end{figure}
As power comparisons with too liberal procedures are difficult to interpret, we excluded the methods $CC_c$, $I_{Pi}$, $S_{Pi}$, and $CC_i$ from our power analysis discussion.
However, their results can be found in the Supplementary Material. 
There, it can, e.g., be observed that the methods $I_{Pi}$, $S_{Pi}$, and $CC_i$ have usually much smaller power than the remaining tests in case of models $M5-M7$.
The power of the remaining procedures $I_{Pc}$, $S_{Pc}$, $I_P$, $S_P$, $I_{PB}$, and $S_{PB}$ depends highly on the considered setting.
This is illustrated in Figure~\ref{fig:BB_m5_scenario2_grid} for Model $M5$ in case of Scenario 2 with BB processes and two error distributions (normal and lognormal). 
We observe that the power is generally higher under lognormal errors than under Gaussian errors.
Additionally, the type of test statistics ($I$ vs. $S$) seem to influence the power. 
For lognormal distributed data, integral-type statistics have a higher power and in case of Gaussian errors, supremum-type test statistics seem to be superior.
Corresponding plots for other combinations of $\{\text{BB}, \text{Zhang}\}$, $\{M2,M3, M5, M6, M7\}$, and $\{\text{Scenario }1, \text{Scenario }2\}$ are presented in the Supplementary Material. 
The results also show that neither the integral- nor the supremum-type statistic is generally better than the other. 
For example, shattered point plots of the power difference between $I_P$ and $S_P$ show an advantage of $S_P$ under Model $M7$ while $I_P$ tends to have greater power under
Models $M5-M6$. 
Therefore, it is impossible to give a clear recommendation for one of these approaches.
Some general patterns are nevertheless visible.
The power increases with higher correlations and settings with Zhang processes lead to a greater power variability.
Furthermore, power also depends strongly on the considered model. 
As expected, power increases as the difference between the two mean functions increases, e.g. when moving from $M2$ to $M3$ or from $M5$ through $M6$ to $M7$.
Finally, if we compare the behaviour of the two resampling methods (permutation vs. parametric bootstrap), the two methods exhibit comparable power in all settings.

\begin{table}[t!]
	\caption{Comparison of the empirical power obtained in 	Models $M2$, $M3$, and $M5$ - $M7$ for Scenario 2 with BB and Zhang processes of the tests $I_{Pc}$ vs. $I_P$, $S_{Pc}$ vs. $S_P$, $I_{Pc}$ vs. $I_{PB}$, and $S_{Pc}$ vs. $S_{PB}$ for selected sample sizes configurations. 
		Numerical values indicate the percentage of simulation runs in which a particular test showed higher (b - better), equal (t - tie), or lower (w - worse) power relative to the corresponding test. 
		The sample sizes are sorted by increasing of missing data amount.}
	\label{tab:P_PS_complete_incomplete}
	\begin{tabularx}{\textwidth}{llRRRRRR}
		\toprule
		& $(n_1,n_2,n_3)$ & \multicolumn{3}{c}{$I_{Pc}$ vs. $I_P$} & \multicolumn{3}{c}{$S_{Pc}$ vs. $S_P$}\\
		Process  & & b & t & w & b & t & w\\
		\midrule
		\multirow{6}{*}{BB} & $(21,5,4)$ & 53.33 & 46.67 & 0.00 &  60.00 & 40.00 &  0.00\\
		& $(18,6,6)$ & 46.67 & 33.33 & 20.00 & 43.33 & 36.67 & 20.00\\
		& $(15,8,7)$ & 23.33 & 23.33 & 53.33 & 13.33 & 26.67 & 60.00\\
		& $(12,9,9)$ & 13.33 & 16.67 & 70.00 & 0.00 & 13.33 & 86.67\\
		& $(9,11,10)$ & 3.33 & 0.00 & 96.67 & 0.00 & 10.00 & 90.00\\
		& $(6,12,12)$ & 0.00 & 3.33 & 96.67 & 0.00 & 0.00 & 100.00\\
		\midrule
		\multirow{6}{*}{Zhang} & $(21,5,4)$ & 67.22 & 26.11 & 6.67 & 64.44 & 34.44 & 1.11\\
		& $(18,6,6)$ & 49.44 & 26.11 & 24.44 & 51.67 & 32.22 & 16.11\\
		& $(15,8,7)$ & 39.44 & 22.78 & 37.78 & 34.44 & 28.89 & 36.67\\
		& $(12,9,9)$ & 25.56 & 18.89 & 55.56 & 21.67 & 25.00 & 53.33\\
		& $(9,11,10)$ & 21.11 & 18.89 & 60.00 & 13.89 & 22.22 & 63.89\\
		& $(6,12,12)$ & 30.56 & 16.67 & 52.78 & 34.44 & 13.89 & 51.67\\
		\midrule
		& $(n_1,n_2,n_3)$ & \multicolumn{3}{c}{$I_{Pc}$ vs. $I_{PB}$} & \multicolumn{3}{c}{$S_{Pc}$ vs. $S_{PB}$}\\
		Process & & b & t & w & b & t & w\\
		\midrule
		\multirow{6}{*}{BB} & $(21,5,4)$  &  66.67 & 33.33 & 0.00 &  63.33 &  36.67 &  0.00\\
		& $(18,6,6)$   & 70.00  & 30.00 & 0.00 & 40.00 & 40.00 & 20.00 \\
		& $(15,8,7)$   & 56.67 & 23.33 & 20.00 & 10.00 & 26.67 & 63.33\\
		& $(12,9,9)$   &  40.00 & 13.33 & 46.67 & 0.00 & 13.33 & 86.67\\
		& $(9,11,10)$  &  23.33 & 3.33 & 73.33 & 0.00 & 10.00 & 90.00 \\
		& $(6,12,12)$  & 10.00  & 0.00 & 90.00 & 0.00 & 0.00 & 100.00\\
		\midrule
		\multirow{6}{*}{Zhang} & $(21,5,4)$  &  79.44 & 20.56 & 0.00 & 65.00 & 33.89 & 1.11\\
		& $(18,6,6)$  & 77.22 & 21.67 & 1.11 & 60.56 & 31.11 & 8.33\\
		& $(15,8,7)$  &  75.56 & 19.44 & 5.00 & 47.78 & 28.33 & 23.89\\
		& $(12,9,9)$  &  66.67 & 17.22 & 16.11 & 28.89 & 24.44 & 46.67\\
		& $(9,11,10)$ &  55.56 & 17.78 & 26.67 & 21.67 & 21.11 & 57.22 \\
		& $(6,12,12)$ & 51.11 & 15.56 & 33.33 & 30.56 & 15.56 & 53.89\\
		\bottomrule
	\end{tabularx}
\end{table}

We next give a deeper inside into the methods' behaviour through pairwise comparisons of certain methods. 
To this end Table~\ref{tab:P_PS_complete_incomplete} displays the percentage of simulation runs in which a particular test showed higher (b - better), equal (t - tie), or lower (w - worse) power relative to the corresponding test in case of Scenario~2.
It is readily seen that the tests $I_{Pc}$ and $S_{Pc}$ based on complete data
are superior to $I_{P}$ and $S_{P}$ in many cases with a small amount of missing data ($(n_1,n_2,n_3)=(21,5,4),(18,6,6)$).
However, in this case many ties are present.
As the proportion of incomplete observation increases ($(n_1,n_2,n_3)=(15,8,7),(12,9,9),(9,11,10),(6,12,12)$), the new methods are more powerful than $I_{Pc}$ and $S_{Pc}$ in most cases.  
Here, $I_{Pc}$ and $S_{Pc}$ are only more powerful in some cases, corresponding to settings where they exhibit a liberal behaviour.
We next compare $I_{Pc}$ vs. $I_{PB}$ and $S_{Pc}$ vs. $S_{PB}$.
For the supremum-type statistic a similar pattern as in the comparison with the permutation approach can be observed for the parametric bootstrap version $S_{PB}$.  
However, for the integral-type statistic, the parametric bootstrap version  $I_{PB}$ seldom has an advantage over $I_{Pc}$. 
In particular, in case of Zhang processes, $I_{Pc}$ outperforms $I_{PB}$ for all sample site settings. 
This is consistent with the often conservative nature of $I_{PB}$. 
Nevertheless, in case of BB processes, $I_{PB}$ has an advantage in case of larger numbers of incomplete pairs ($n_2,n_3\geq 9$).

\begin{figure}[!h]
	\centering
	\includegraphics[width=\textwidth]{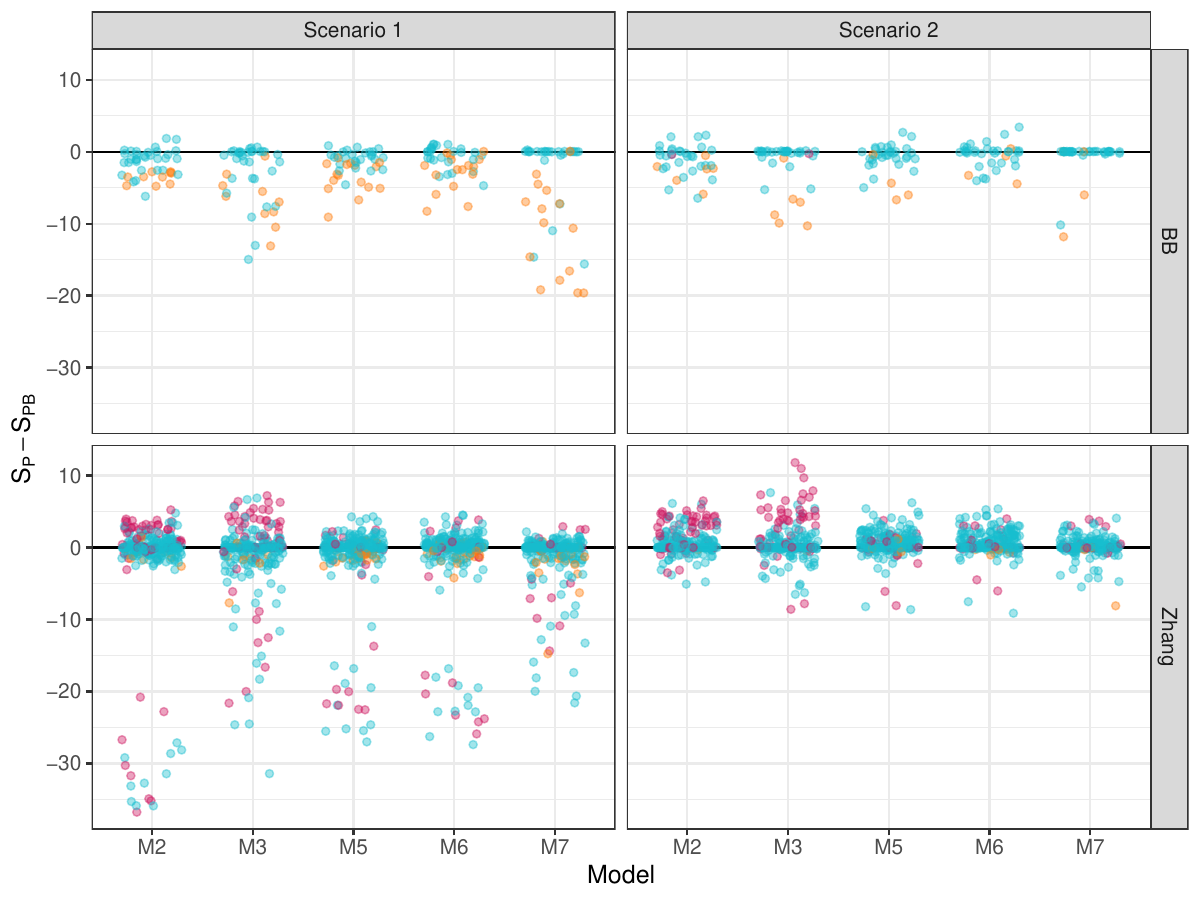}
	\caption{Shattered point plots for the difference in empirical powers (as percentages) of methods $S_P$ and $S_{PB}$. Turquoise points indicate cases, when $S_{PB}$ controls the type I error rate (empirical value inside the binomial interval $[3.6,6.4]$), pink indicates a conservative and orange a liberal behaviour.}
	\label{fig:I_P.I_PB}
\end{figure}

The comparison of the two resampling approaches is of special interest as only the parametric bootstrap versions are generally asymptotic valid.
For the integral-type statistic, $I_P$ has a power advantage due to the often too conservative behaviour of $I_{PB}$. 
Only judging from the simulation results, $I_P$ seems to be the preferred method. 
The comparison of the methods $S_P$ and $S_{PB}$ is more balanced as $S_{PB}$ is not as conservative as $I_{PB}$. 
Figure~\ref{fig:I_P.I_PB} shows shattered point plots of the power difference between $S_{P}$ and $S_{PB}$. 
It shows a certain power advantage for $S_P$ in case of Models $M5-M6$ (especially for BB processes) while $S_{PB}$ exhibits better power for the Models $M2$ and $M7$ as well as M3 in case of Zhang processes. 
A detailed inspection of these difference show that they are likely due the respective conservative/liberal type I error control. 
A corresponding comparison of $I_P$ and $I_{PB}$ can be found in the Supplementary Material. 

To sum up, no test is generally superior to the others. 
This holds for the statistics types ($I$ vs. $S$) as well as the chosen resampling approaches ($P$ vs. $PB$).
The procedures $I_P, S_P$, and $S_{PB}$ benefit the most from the inclusion of incomplete data, especially in case of small numbers of complete pairs. 
In contrast, the conservative behaviour of $I_{PB}$ reduces its power in several settings. 
Consequently, $I_P$, $S_P$, and $S_{PB}$ can be recommended based on our simulation results.

\section{Data Analysis of NO$_2$ Emission} \label{sec:real_data_example}
We illustrate the proposed methods using functional measurements of annual average Nitrogen Dioxide NO$_2$ concentrations in U.S. cities. 
A previous study by \textcite{clark2017changes} discovered a substantial decline from 2000 to 2010. 
The decline has been attributed legal regulations and technological progress aimed at reducing toxic air pollution and improving public health by reducing smog and acid rain. The objective of this section is not only to confirm this finding via \textsc{fda}, but also to illustrate how our proposed procedures combine information from complete and incomplete pairs.
We would also like to illustrate the use of the tests considered in this paper and show some differences between them.
The related \textsf{R}-scripts can be found in the Supplementary Material \parencite{baumeister_supplementary_2026}.

\begin{figure}[!t]
	\centering
	\includegraphics[width=0.97\textwidth, height=0.6\textheight]{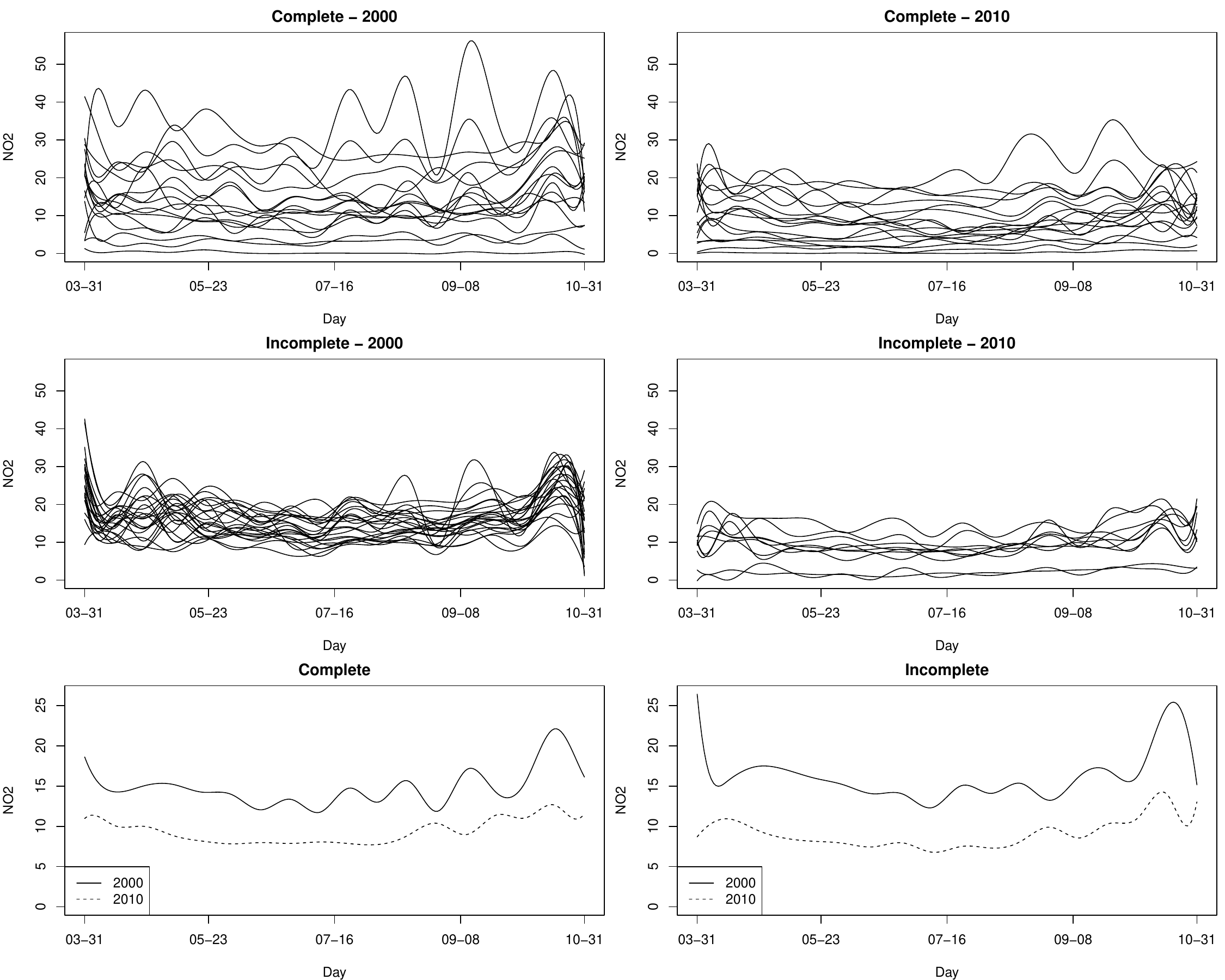}
	\caption{Trajectories for functional observations for complete (first row), incomplete (second row) pairs, and the sample mean functions (last row) for the full data set.}
	\label{fig_rde_1}
\end{figure}
\begin{figure}[!t]
	\centering
	\includegraphics[width=0.97\textwidth, height=0.6\textheight]{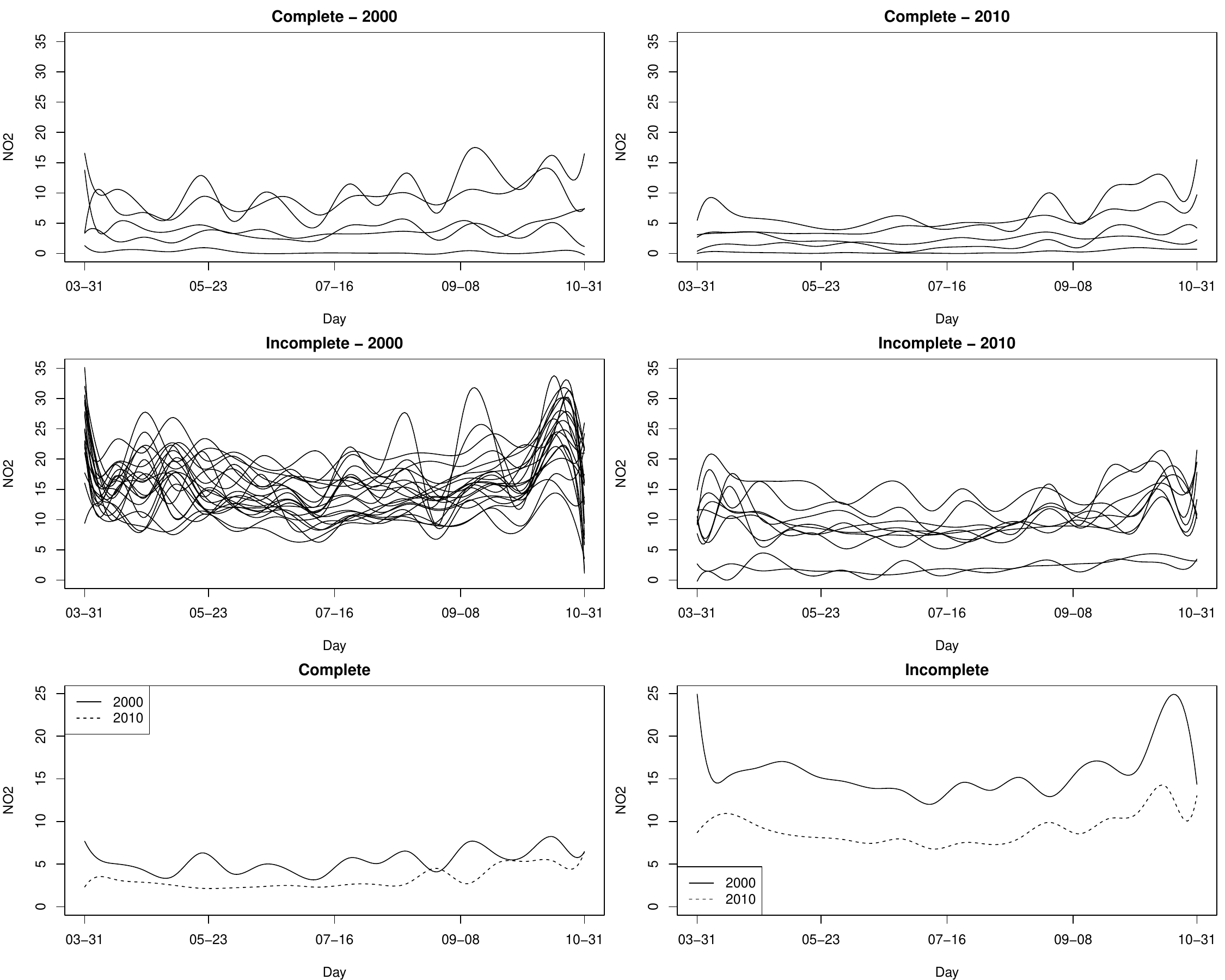}
	\caption{Trajectories for functional observations for complete (first row), incomplete (second row) pairs, and the sample mean functions (last row) for the selected data set.}
	\label{fig_rde_2}
\end{figure}

The daily NO$_2$ concentration measurements were obtained from the U.S. air pollution data set available at \url{https://www.kaggle.com/datasets/sogun3/uspollution}, which was also studied in \textcite{Zhu_et_al_2022}.
We consider data from 2000 and 2010 over the period from 31 March to 31 October. 
This interval was selected as there were much more missing measurements at the beginning and end of these years. 
Cities that had a large remaining missingness during this period were excluded from our analysis. 
Following \textcite{Zhu_et_al_2022}, the daily concentration measurement raw curves for each year and city were smoothed using a B-spline basis of order $4$ and with $20$ basis functions. 
The resulting curves were evaluated over a common grid of time points $t\in\{1,\dots,215\}$ corresponding to $215$ days between 31 March to 31 October.
In our analysis, the year 2000 defines the first group and the year 2010 the second group, with cities regarded as independent observation units.
For cities observed in both years, we have matched pairs on functional data. 
Other cities were only observed for one of the two years, leading to incomplete functional observations.
Therefore, the data has the incomplete paired structure and can be represented by Model~(\ref{eq:missingdata}).

Let $m_{2000}(t)$ and $m_{2010}(t)$ denote the point-wise means in year 2000 and 2010, respectively ($t\in[1, 215]$). 
We are then interested in the null hypothesis
\begin{align}\label{eq:hypothesis_data_example}
	\mathcal{H}_0:\, \forall t\in[1, 215]:\, m_{2000}(t) = m_{2010}(t).
\end{align}

Overall, there are $n_1=17$ complete pairs with measurements for both years, $n_2=23$ functional observations for the year 2000 only, and $n_3=10$ observations for the year 2010 only. 
We refer to this sample as the \textit{full} data set. 
The concrete list of cities for each observation category is given in \texttt{additional.pdf} in the Supplementary Material \parencite{baumeister_supplementary_2026}.  

To illustrate the behaviour of the methods under a higher number of missing functions, we also consider a \textit{selected} data set. 
Here, observations with very large concentration levels were removed as follows: from the complete pairs, we removed all pairs containing values greater than 20, while from the observations available only for 2000, we removed curves containing values greater than 40.
This results in $n_1=5$ complete pairs and $n_2=21$ functional observations available only for 2000. 
The $n_3=10$ observations available only for 2010 were kept and are the same as before. 
The observed trajectories and sample mean functions are presented in Figures~\ref{fig_rde_1} and~\ref{fig_rde_2} for the full and selected data set, respectively.

\begin{table}[!t]
	\caption{Collected $p$-values, empirical type I error rate, and power (as percentages) of all tests for the NO$_2$ emission data sets. Column DS denotes type of data set (\textit{full} (f) or \textit{selected} (s)). The empirical type I error rates are given in bold when they are greater than $6.4\%$ (i.e., the upper limit of the $95\%$ binomial confidence interval \parencite[cf.][]{DuchesneFrancq2015}), indicating that the test is too liberal. The corresponding $p$-values and powers of these tests are also given in bold. The $p$-values are calculated via Equations~(\ref{eq:perm_pvalues}) and (\ref{eq:boot_pvalues}). }\label{tab:results_data_analysis}
	\centering
	\begin{tabular}{@{}clrrrrrrrrrr@{}}
		\toprule
		Distr. & DS & $I_{Pc}$ & $S_{Pc}$ & $CC_c$ & $I_{Pi}$ & $S_{Pi}$ & $CC_i$ & $I_P$ & $S_{P}$ & $I_{PB}$ & $S_{PB}$ \\
		\midrule
		\multicolumn{12}{l}{$p$-values}\\
		& f & $<0.1$ & $<0.1$ & $\textbf{<0.1}$ & $\textbf{<0.1}$ & $\textbf{<0.1}$ & $\textbf{<0.1}$ & $<0.1$ & $<0.1$ & $<0.1$ & $<0.1$ \\
		& s & $\textbf{<0.1}$ & 8.2 & $\textbf{<0.1}$ & $\textbf{<0.1}$ & $\textbf{<0.1}$ &$\textbf{<0.1}$ & $<0.1$ & $<0.1$ & 0.2 & 1.8 \\
		\midrule
		\multicolumn{12}{l}{Empirical Type I Error Rate} \\
		\multirow{2}{*}{$\mathcal{N}$} & f &  5.3&4.0&\textbf{16.9}&\textbf{63.7}&\textbf{65.6}&\textbf{68.8}&5.1&4.1&5.1&3.8\\
		& s & \textbf{6.5}&6.1&\textbf{48.7}&\textbf{52.5}&\textbf{56.0}&\textbf{61.2}&4.6&4.3&4.6&5.8\\
		\multirow{2}{*}{$t_5$} & f & 5.2 & 5.3&\textbf{16.2}&\textbf{58.5}&\textbf{61.5}&\textbf{64.4}&5.9&5.2&5.4&4.5\\
		& s & \textbf{6.7}&6.2&\textbf{45.9}&\textbf{51.3}&\textbf{55.0}&\textbf{59.4}&6.1&5.0&4.3&4.6\\
		\multirow{2}{*}{$\chi^2_{10}$} & f & 5.2&5.3&\textbf{17.8}&\textbf{61.9}&\textbf{64.3}&\textbf{69.1}&4.8&4.5&4.7&5.3\\
		& s & \textbf{7.7}&\textbf{7.9}&\textbf{50.5}&\textbf{47.9}&\textbf{52.3}&\textbf{56.8}&5.7&6.2&6.2&\textbf{6.8}\\
		\midrule
		\multicolumn{12}{l}{Empirical Power} \\
		\multirow{2}{*}{$\mathcal{N}$} & f & 100& 100& \textbf{100}& \textbf{99.9}& \textbf{99.9}&\textbf{100}& 100 &100&100&100\\
		& s & \textbf{63.6} & 45.6& \textbf{99.5}& \textbf{99.5} &\textbf{99.7}& \textbf{99.7}& 98.4 & 86.4& 99.9& 99.4\\
		\multirow{2}{*}{$t_5$} &	f & 100&100&\textbf{100}&\textbf{99.6}&\textbf{99.8}&\textbf{99.8}&100&100&100&100\\
		& s & \textbf{69.2}&54.0&\textbf{99.1}&\textbf{99.3}&\textbf{99.8}&\textbf{99.9}&98.3&88.7&99.2&99.6\\
		\multirow{2}{*}{$\chi^2_{10}$} & f & 100&100&\textbf{100}&\textbf{99.7}&\textbf{100}&\textbf{100}&100&100&100&100\\
		& s & \textbf{66.5}&\textbf{47.4}&\textbf{99.6}&\textbf{99.6}&\textbf{99.9}&\textbf{99.9}&99.1&90.4&99.8&\textbf{99.5}\\
		\bottomrule
	\end{tabular}
\end{table}
The resulting $p$-values for testing the null hypothesis~(\ref{eq:hypothesis_data_example}) are given in Table~\ref{tab:results_data_analysis}. 
For the full data set, we observe that all considered tests reject the null hypothesis. 
Together with the sample mean functions displayed in Figure~\ref{fig_rde_1}, this indicates that the NO$_2$ concentrations significantly decreased from 2000 to 2010, confirming the results of \textcite{clark2017changes}.
For the selected data set, all procedures except $S_{Pc}$ reject the null hypothesis. 
The difference is plausible due to the smaller number of complete pairs and the reduced separation between the estimated mean functions. 

To shed more light on the behaviour of the different tests in these settings, we conduct additional Plasmode-type simulation studies based on the considered data sets. 
To mimic the structure of the full and selected data sets, we simulated a sample with $n_1$ complete functional pairs and two samples with $n_2$ and $n_3$ independent univariate functional observations. 
We set $(n_1,n_2,n_3) = (17,23,10)$ for the full and $(n_1,n_2,n_3) =(5,21,10)$ for the selected data setting, respectively.
The covariance functions were chosen equally to the sample covariance functions from the subsets of the original data, i.e., for the first sample of complete pairs, the covariance function is equal to the sample covariance function for complete data, while for the incomplete samples, we used the sample covariance functions estimated separately for each independent sample.
To investigate type I error control, each functional observation was generated with a common mean function equal to the sample mean function of the pooled sample containing all $2\cdot n_1+n_2+n_3$ observed curves.
To study the power, the mean function for each of the three samples was set equal
to the corresponding empirical mean function estimated from the corresponding original subset.
The data were simulated observations were generated using standard Gaussian, student $t_5$ with five degrees of freedom, and $\chi_{10}^2$-distributions with ten degrees of freedom.
In each case, the expected value and covariance functions were equal to the sample mean functions and sample covariance functions, respectively.

The results of this simulations study are presented in Table~\ref{tab:results_data_analysis}. 
Similar to the results from Section~\ref{sec:simulation}, the methods $CC_c$, $I_{Pi}$, $S_{Pi}$, and $CC_i$ are too liberal for both data sets. 
Therefore, their results may be misleading and should not be trusted in practice. 
Fortunately, for the full data set, the remaining procedures keep the assigned type I error level and have power equal to or close to $100\%$, which suggests the correctness of their results in this setting.
On the other hand, for the selected data set, the method $I_{Pc}$ is slightly too liberal, likely due to a smaller number of paired observations ($n_1=5$). 
Therefore, its results must be considered with some caution. 
Moreover, it is less powerful than the newly proposed tests, which all control the type I error level well and can recommend for this setting as well. 
We also observe that $S_{Pc}$ has smaller power than $I_{Pc}$, which could explain the non rejection in the illustrative analysis.

To sum up, in the case of the selected data set, where the number of complete pairs is small, the proposed tests benefit from incorporating the additional information from the incomplete observations. 
This result is in line with the direct comparison of the new methods with $I_{Pc}$ and $S_{Pc}$, compare Table~\ref{tab:P_PS_complete_incomplete} in Section~\ref{sec:simulation}.

\section{Discussion}\label{sec:end}
We have developed hypothesis tests based on an integral- and supremum-type statistic for incompletely paired functional data that combine information from complete pairs and observations available in only one of the two groups. 
To this end, and to ensure accurate finite-sample performance, we have considered both, a parametric bootstrap and a permutation procedure. 
Both resampling approaches can be viewed as generalizations of the procedures proposed by \textcite{smaga_note_2020} for completely paired functional data whose theoretical validities had not been established so far.

Our mathematical analyses reveal an important difference between the two resampling approaches.
The parametric bootstrap tests are asymptotically valid under the general incomplete paired functional data settings. 
In contrast, the permutation tests are not asymptotically correct in general. 
However, their validity can be established in case of homoscedastic covariance functions, balanced missingness, or the complete-pairs setting studied by \textcite{smaga_note_2020}. 
As a by-product, this establishes the previously lacking theoretical validity of the resampling tests proposed by \textcite{smaga_note_2020}.

We have investigated all four proposed procedures, the two permutation tests \(I_P\) and \(S_P\) as well as the two parametric bootstrap tests \(I_{PB}\) and \(S_{PB}\), together with existing competitors in extensive simulation studies. 
Despite the theoretical limitation of the permutation tests, they exhibit accurate type I error control in our simulation study in most settings considered, even in cases where the method is not asymptotically valid. 
A possible explanation is that the permutation procedure remains point-wise valid \parencite{amro_permuting_2017}, which seems to contribute to its favourable finite-sample behaviour. 
Overall, the simulation results favour the procedures \(I_P\), \(S_P\), and \(S_{PB}\) for testing equality of mean functions in incompletely paired functional data. 
The integral-based parametric bootstrap approach \(I_{PB}\), though asymptotically valid, appears to be less attractive due its rather conservative behaviour in many settings. 

Our simulation study further suggests that incorporating incomplete observations can lead to substantial power gains compared to complete-case procedures, particularly when the number of complete pairs is small. 
The proposed tests can also outperform the imputation approach of \textcite{jang_bayesian_2021} in the settings considered, especially if the amount of incomplete functions is large. 
The complete-case methods \(CC_c\) and \(CC_i\) of \textcite{crainiceanu2012bootstrap} show the expected liberal behaviour for small sample sizes in several simulation scenarios, and our methods were able to generally outperform them.
The illustrating data analysis of the NO$_2$ concentration data additionally highlights the practical relevance of methods that combine information from complete and incomplete functional observations, particularly in cases with only a few complete pairs.

For the future, we envision several extensions of the proposed methodology. 
One direction could be the investigation, if our way of handling missing functions also work for more than two groups, e.g., in the frameworks of \textcite{cuevas2004anova, martinez2011repeated, zhang_new_2019, munko_general_2025} or in a multivariate case, e.g., in \textcite{gorecki_multivariate_2017, Zhu_et_al_2022, munko_multiple_2025, qiu_global_2026}.
Moreover, the proposed methods are currently only applicable in case of incomplete functional observations in both components. 
An extension to settings in which incomplete observations occur in only one group could be based on adapting the method of \textcite{amro_asymptotic-based_2021} to the functional framework.
Another line of future research could be the development of mathematically valid Rubin-type rules for multiply imputed functional data for our asymptotically non-pivotal test statistics. 
Finally, we plan a software implementation of the proposed methods to facilitate practical applications.
Since then, the \textsf{R}-code of our analyses can be found in the Supplementary Material \parencite{baumeister_supplementary_2026} and can be used for own analyses.

\section*{Acknowledgements}
The work of M. Baumeister and M. Pauly were supported by the German Research Foundation project, Grant no. 565106208.
The work of Ł. Smaga was funded by the National Science Centre, Poland under the Weave-UNISONO (grant no. 2025/07/Y/ST6/00001). 
A part of calculations for the simulation studies was made at the Pozna{\'n} Supercomputing and Networking Center (grant no. Grant pl0253-04).
We thank our student assistant Antonia Köhr for her help with programming and plotting in \textsf{R}.

\section*{CRediT Authorship Contribution Statement}
\textbf{Marl\'{e}ne Baumeister:} Conceptualization, Methodology, Validation, Formal analysis, Data Curation, Writing - Original Draft, Visualization, Project administration;
\textbf{Lubna Amro:} Conceptualization, Software, Writing - Review \& Editing;
\textbf{\L ukasz Smaga:} Conceptualization, Software, Formal analysis, Data Curation, Writing - Original Draft, Visualization, Supervision, Funding acquisition;
\textbf{Markus Pauly:} Conceptualization, Methodology, Validation, Writing - Review \& Editing, Supervision, Funding acquisition.

\section*{Data Availability Statement}
The data that support the findings of this work, e.g., all simulation results, simulation scripts and the scripts for the data analysis are openly available in TUDOdata at \url{https://doi.org/10.17877/TUDODATA-2026-MR0D9JVC}.

\section*{Declaration of generative AI and AI-assisted technologies in the manuscript preparation process.}
During the preparation of this work the authors used \textsc{o}pen\textsc{ai}’s \textsc{c}hat\textsc{gpt} in order to refine phrasing and improve readability. 
After using this tool, the authors reviewed and edited the content as needed and takes full responsibility for the content of the published article.

\printbibliography

\newpage
\appendix
\section{Proofs and Mathematical Theory}\label{sec:proofs}
In this section, some aspects of empirical process theory that are useful to understand the mathematical modelling in this paper, and the technical proofs of all stated theorems are presented.

\subsection{Modelling Functional Data with Empirical Processes}\label{sec:eptheory}
In line with \textcite{dette_functional_2020, munko_multiple_2025}, we handle functional data as realizations of stochastic processes valued in $C^p(\mathcal{T})$ equipped with the norm $\|f\|_\infty=\sup_{t \in \mathcal{T}}|f(t)|$.
This allows us to use the results of Section 1.5 in \textcite{van_der_vaart_weak_2023} about weak convergence in the space $\ell^{\infty}(\mathcal{T})=\{f:\mathcal{T} \to\br\vert\, \|f\|_\infty<\infty\}$ of bounded functions on $\mathcal{T}$.
This possibility is explained in Example 1.5.1 in \textcite{van_der_vaart_weak_2023} and \textcite{dette_functional_2020} note in their supplement that the results of Section 1.5 in \textcite{van_der_vaart_weak_2023} also hold if the space $\ell^{\infty}(\mathcal{T})$ is replaced by $C^p(\mathcal{T})$.

The following theorem is used for the proof of Theorem 2 , and it is a special case of Lemma 5 in \textcite{munko_multiple_2025}.
They proved a version for an arbitrary dimension (we set $p=2$ in their notation) under a null hypothesis and considered concepts of  Section 1.5  in \textcite{van_der_vaart_weak_2023} and \textcite{dette_functional_2020}.
\begin{satz}\label{thm:clt}
	Under \ref{ass:A1} and \ref{ass:A5}, it holds
	\begin{align}\label{eq:clt}
		\left(\sqrt{n}(\boldsymbol{\hat\eta}-\boldsymbol{\eta})(t)\right)_{t\in\mathcal{T}}\stackrel{d}{\longrightarrow}\mathbf{Y}=(Y_1,Y_2)^\top\sim \GP_2(\mathbf{0},\boldsymbol{\Gamma})
	\end{align}
	for $n\to\infty$ in $C^2(\mathcal{T})$.
\end{satz}
The class of measurable functions $\mathcal{F}=\{f_t:C^2(\mathcal{T})\to\br|f_t(\mathbf{x}_j)=\mathbf{x}_j(t),\,t\in\mathcal{T},\,j\in\{1,\dots,n\}\}$ defines the realizations of the stochastic process $\mathbf{X}$ in Equation~(\ref{eq:data}) in context of the empirical process theory in Section 2 of \textcite{van_der_vaart_weak_2023}.
Because of this, $C^2(\mathcal{T})$ can be associated with $C^2(\mathcal{F})$, compare Example 2.1.3 in \textcite{van_der_vaart_weak_2023}.
In fact, Theorem~\ref{thm:clt} ensures that the class $\mathcal{F}$ is Donsker \parencite[cf.][Equation~2.1.1, p.130]{van_der_vaart_weak_2023}.
That is why empirical process theory can be applied to prove the permutation techniques \parencite[cf.][Section 3.8]{van_der_vaart_weak_2023}.
To ensure a deterministic convergence, we always consider the convergent subsequences of $n_i/(n_2+n_3)$, $i\in\{1,2,3\}$.

\subsection{Proof of Theorem~\ref{thm:distribution_test_statistics}} \label{sec:proof_distribution_test_statistics}
For easier handling, we always set $\mathbf{h}=(1,-1)^\top$.
In general the asymptotic behaviour of the two pointwise test statistics can be concluded from Theorem~\ref{thm:clt}.
Firstly, we infer the pointwise paired t-type statistic $T_{n_1}(\mathbf{x}^{(c)}, \boldsymbol{\hat\Gamma}^{(c)})(t)$, $t\in\mathcal{T}$.
Then, under the null hypothesis $\mathcal{H}_0$ in Equation~(\ref{eq:hypothesis}) and with the Continuous Mapping Theorem, it follows from Theorem~\ref{thm:clt} that
\begin{align}\label{eq:diffclt}
	\left(\sqrt{n_1}(\hat\eta^{(c)}_1-\hat\eta^{(c)}_2)(t)\right)_{t\in\mathcal{T}}\stackrel{d}{\longrightarrow}Y_{1-2}\sim \GP(0,\gamma_{1-2})
\end{align}
for $n_1\to\infty$ in $C(\mathcal{T})$.
The Equation~(\ref{eq:diffclt}) is a specific case of the assertion of Lemma 5 in \textcite{munko_multiple_2025}.
From the uniform consistency of $\boldsymbol{\hat\Gamma}$ for $\boldsymbol{\Gamma}$ (Lemma 6 of \textcite{munko_multiple_2025}), it follows that $\mathbf{h}^\top\boldsymbol{\hat\Gamma}^{(c)}\mathbf{h}=\hat\gamma_{1-2}$ is consistent for $\gamma_{1-2}$.

Secondly, to infer the pointwise Welch-type statistic $T_{n_2+n_3}(\mathbf{x}^{(m)}, \boldsymbol{\hat\Gamma}_0^{(m)} )(t)$, $t\in\mathcal{T}$, a component-wise convergence in distribution can be concluded from Theorem~\ref{thm:clt}, because of the missing values in $\mathbf{x}^{(m)}$ and the assumption of independent observations in general.
With Assumption \ref{ass:A4}, it follows that
\begin{align*}
	&\left(\sqrt{\frac{n_2n_3}{n_2+n_3}}(\hat\eta^{(m)}_1-\eta^{(m)}_1)(t)\right)_{t\in\mathcal{T}}\\
	=&\left(\sqrt{\frac{n_3}{n_2+n_3}}\sqrt{n_2}(\hat\eta^{(m)}_1-\eta^{(m)}_1)(t)\right)_{t\in\mathcal{T}}\stackrel{d}{\longrightarrow}\sqrt{\kappa_3}\,Y_1\sim\GP\left(0,\kappa_3\gamma_1\right),\\
	&\left(\sqrt{\frac{n_2n_3}{n_2+n_3}}(\hat\eta^{(m)}_2-\eta^{(m)}_2)(t)\right)_{t\in\mathcal{T}}\\
	=&\left(\sqrt{\frac{n_2}{n_2+n_3}}\sqrt{n_3}(\hat\eta^{(m)}_2-\eta^{(m)}_2)(t)\right)_{t\in\mathcal{T}}\stackrel{d}{\longrightarrow}\sqrt{\kappa_2}\,Y_2\sim\GP\left(0,\kappa_2\gamma_2\right)
\end{align*}
for $\min\{n_2,n_3\}\to\infty$ in $C(\mathcal{T})$. 
Under the null hypothesis $\mathcal{H}_0$ in Equation~(\ref{eq:hypothesis}), because of the independence of the two groups in $\mathbf{x}^{(m)}$, and with the Continuous Mapping Theorem, it follows that
\begin{align}\label{eq:welchclt}
	\left(\sqrt{\frac{n_2n_3}{n_2+n_3}}(\hat\eta^{(m)}_1-\hat\eta^{(m)}_2)(t)\right)_{t\in\mathcal{T}}\stackrel{d}{\longrightarrow}\sqrt{\kappa_3}\,Y_1-\sqrt{\kappa_2}\,Y_2\sim\GP\left(0,\kappa_3\gamma_1+\kappa_2\gamma_2\right)
\end{align}
for $\min\{n_2,n_3\}\to\infty$ in $C(\mathcal{T})$.
It holds
\begin{align*}
	\mathbf{h}^\top\boldsymbol{\hat\Gamma}_0^{(m)}(t,t)\mathbf{h}=\frac{n_3}{n_2+n_3}\hat\gamma_1^{(m)}(t,t)+\frac{n_2}{n_2+n_3}\hat\gamma_2^{(m)}(t,t)
\end{align*}
for every $t\in\mathcal{T}$.
From the uniform consistency of $\boldsymbol{\hat\Gamma}$ for $\boldsymbol{\Gamma}$ \parencite[Lemma 6 of][]{munko_multiple_2025} and with Assumption~\ref{ass:A4}, it follows that $\mathbf{h}^\top\boldsymbol{\hat\Gamma}_0^{(m)}\mathbf{h}=n_3/(n_2+n_3)\hat\gamma_1^{(m)}+n_2/(n_2+n_3)\hat\gamma_2^{(m)}$ is consistent for $\mathbf{h}^\top\boldsymbol{\Gamma}_0\mathbf{h}:=\kappa_3\gamma_1+\kappa_2\gamma_2$.
As the limiting variance is uniformly bounded away from zero, it follows from the consistency of both variance estimators, Equations~(\ref{eq:diffclt}) and (\ref{eq:welchclt}), and Slutzky's Theorem that 
\begin{align}\label{eq:asymp_distr_test_statistics}
	\left(T_{n_1}(\mathbf{x}^{(c)},\boldsymbol{\hat\Gamma}^{(c)})(t),\,T_{n_2+n_3}(\mathbf{x}^{(m)}, \boldsymbol{\hat \Gamma}_0^{(m)})(t) \right)_{t\in\mathcal{T}}\stackrel{d}{\longrightarrow}(Z_1,Z_2)
\end{align}
in $C(\mathcal{T})^2$, where $Z_1$ and $Z_2$ are stochastically independent because of the independence of the data parts $\mathbf{x}^{(c)}$ and $\mathbf{x}^{(m)}$  and it holds
\begin{align*}
	&\cov(Z_1(t),Z_1(s))=\cov\left(\frac{Y_{1-2}(t)}{\sqrt{\gamma_{1-2}(t,t)}},\, \frac{Y_{1-2}(s)}{\sqrt{\gamma_{1-2}(s,s)}}  \right)\\
	=\,& \sqrt{\gamma_{1-2}(t,t)\gamma_{1-2}(s,s)}^{-1}\cov\left(Y_{1-2}(t),Y_{1-2}(s)\right)\\
	=\,& \frac{\gamma_{1-2}(s,t)}{\sqrt{\gamma_{1-2}(s,s)\gamma_{1-2}(t,t)}}
\end{align*}
and 
\begin{align*}
	\cov(Z_2(t),Z_2(s))=\,&\cov\left(\frac{\sqrt{\kappa_3}\,Y_1(t)-\sqrt{\kappa_2}\,Y_2(t)}{\sqrt{\kappa_3\gamma_1(t,t)+\kappa_2\gamma_2(t,t)}},\, \frac{\sqrt{\kappa_3}\,Y_1(s)-\sqrt{\kappa_2}\,Y_2(s)}{\sqrt{\kappa_3\gamma_1(s,s)+\kappa_2\gamma_2(s,s)}}  \right)\\
	=\,& \frac{\cov\left(\sqrt{\kappa_3}\,Y_1(t)-\sqrt{\kappa_2}\,Y_2(t),\sqrt{\kappa_3}\,Y_1(s)-\sqrt{\kappa_2}\,Y_2(s)\right)}{\sqrt{\kappa_3\gamma_1(t,t)+\kappa_2\gamma_2(t,t)}\sqrt{\kappa_3\gamma_1(s,s)+\kappa_2\gamma_2(s,s)}}\\
	=\,& \frac{\kappa_3\gamma_1(s,t)+\kappa_2\gamma_2(s,t)}{\sqrt{\kappa_3\gamma_1(t,t)+\kappa_2\gamma_2(t,t)}\sqrt{\kappa_3\gamma_1(s,s)+\kappa_2\gamma_2(s,s)}}
\end{align*}
for all $s,t\in\mathcal{T}$.
For both $Z_1$ and $Z_2$, it holds $\var(Z_i(t))=1$, $i\in\{1,2\}$, $t\in\mathcal{T}$.
Similarly to the weighted test statistic in \textcite{amro_permuting_2017}, because of the independence of the test statistics $T_{n_1}(\mathbf{x}^{(c)}, \boldsymbol{\hat\Gamma}^{(c)})(t)$ and $T_{n_2+n_3}(\mathbf{x}^{(m)}, \boldsymbol{\hat\Gamma}_0^{(m)} )(t)$ for every $t\in\mathcal{T}$, we can argue with $a_n\to a$ that 
\begin{align}
	\left(K_n(\mathbf{x}, t)\right)_{t\in\mathcal{T}}\stackrel{d}{\longrightarrow}\sqrt{a}Z_1^2+\sqrt{1-a}Z_2^2
\end{align}
in $C(\mathcal{T})$.
The first assertion follows with the continuity of the functions  $t\mapsto\int_\mathcal{T}\,\cdot\, dt$ and $t\mapsto\sup_{t\in\mathcal{T}}(\,\cdot\,)$ on $C(\mathcal{T})$.

To prove the second assertion, we argue that we have the following convergences in probability:
\begin{align*}
	\frac{1}{n_1}\,\left(T^2_{n_1}\left(\mathbf{x}^{(c)},\boldsymbol{\hat\Gamma}\right)(t)\right)_{t\in\mathcal{T}}
	&=\left(\frac{\left(\mathbf{h}^\top\boldsymbol{\hat{\eta}}^{(c)}(t)\right)^2}{\mathbf{h}^\top\boldsymbol{\hat{\Gamma}}^{(c)}(t,t)\mathbf{h}}\right)_{t\in\mathcal{T}}
	\stackrel{p}{\longrightarrow}\left(\frac{\left(\mathbf{h}^\top\boldsymbol{\eta}(t)\right)^2}{\mathbf{h}^\top\boldsymbol{\Gamma}(t,t)\mathbf{h}}\right)_{t\in\mathcal{T}},\\
	\frac{n_2+n_3}{n_2n_3}\,\left(T^2_{n_2+n_3}\left(\mathbf{x}^{(m)},\boldsymbol{\hat\Gamma}_0\right)(t)\right)_{t\in\mathcal{T}}
	&=	\left(\frac{\left(\mathbf{h}^\top\boldsymbol{\hat{\eta}}^{(m)}(t)\right)^2}{\mathbf{h}^\top\boldsymbol{\hat{\Gamma}}_0^{(m)}(t,t)\mathbf{h}}\right)_{t\in\mathcal{T}}
	\stackrel{p}{\longrightarrow}\left(\frac{\left(\mathbf{h}^\top\boldsymbol{\eta}(t)\right)^2}{\mathbf{h}^\top\boldsymbol{\Gamma}_0(t,t)\mathbf{h}}\right)_{t\in\mathcal{T}}.
\end{align*}
With $a_n\to a$ and the Continuous Mapping Theorem, it follows that
\begin{align*}
	&\left(\sqrt{a_n}\;\frac{\left(\mathbf{h}^\top\boldsymbol{\hat{\eta}}^{(c)}(t)\right)^2}{\mathbf{h}^\top\boldsymbol{\hat{\Gamma}}^{(c)}(t,t)\mathbf{h}}+\sqrt{1-a_n}\;\frac{\left(\mathbf{h}^\top\boldsymbol{\hat{\eta}}^{(m)}(t)\right)^2}{\mathbf{h}^\top\boldsymbol{\hat{\Gamma}}_0^{(m)}(t,t)\mathbf{h}}\right)_{t\in\mathcal{T}}\\
	\stackrel{p}{\longrightarrow}&\left(\sqrt{a}\;\frac{\left(\mathbf{h}^\top\boldsymbol{\eta}(t)\right)^2}{\mathbf{h}^\top\boldsymbol{\Gamma}(t,t)\mathbf{h}}+\sqrt{1-a}\;\frac{\left(\mathbf{h}^\top\boldsymbol{\eta}(t)\right)^2}{\mathbf{h}^\top\boldsymbol{\Gamma}_0(t,t)\mathbf{h}}\right)_{t\in\mathcal{T}}:=\left(\Xi(t)\right)_{t\in\mathcal{T}}.
\end{align*}
The Continuous Mapping Theorem allows applying the functions $t\mapsto\int_\mathcal{T}\,\cdot\, dt$ and $t\mapsto\sup_{t\in\mathcal{T}}(\,\cdot\,)$ on the latter convergence.
If the alternative $\eta_1(t)\not=\eta_2(t)$ holds for some $t\in\mathcal{T}$, the alternative holds also in a neighbourhood $\mathcal{S}$ of $t$, because of the continuity of $t\mapsto\eta(t)$.
Then, it holds $\Xi(\xi)>0$ for every $\xi\in S$.
For the supremum-type statistic, this implies: $\sup_{\xi\in\mathcal{S}}\Xi(\xi)>0$.
As we assume $\mathcal{T}$ to be a compact subset of $\br$ while considering the integral-type statistics, $\mathcal{S}$ is also a compact subset of $\br$, and therefore
and $\int_{\mathcal{S}}\Xi(\xi)d\xi>0$.
From these two observations, it follows the second assertion.

\subsection{Proof of Theorem~\ref{thm:distribution_perm_test_statistic}}\label{sec:proof_distribution_perm_test_statistics}
To prove these statements, it has to be argued that the processes of the permutation test statistics $\{T_{n_1}(\mathbf{x}_\tau^{(c)},\boldsymbol{\hat\Gamma}_\tau^{(c)})(t)|\,t\in\mathcal{T}\}$ and $\{T_{n_2+n_3}(\mathbf{x}_\pi^{(m)},\boldsymbol{\hat\Gamma}_{0,\pi}^{(m)})(t)|\,t\in\mathcal{T}\}$,
\begin{align*}
	T_{n_1}\left(\mathbf{x}_\tau^{(c)},\boldsymbol{\hat\Gamma}_\tau^{(c)}\right)(t)=\sqrt{n_1}\frac{\mathbf{h}^\top\boldsymbol{\hat{\eta}}_\tau^{(c)}(t)}{\sqrt{\mathbf{h}^\top\boldsymbol{\hat{\Gamma}}_\tau^{(c)}(t,t)\mathbf{h}}},\\
	T_{n_2+n_3}\left(\mathbf{x}_{0,\pi}^{(m)},\boldsymbol{\hat\Gamma}_\pi^{(m)}\right)(t)=\sqrt{n_2+n_3}\frac{\mathbf{h}^\top\boldsymbol{\hat{\eta}}_\pi^{(m)}(t)}{\sqrt{\mathbf{h}^\top\boldsymbol{\hat{\Gamma}}_{0,\pi}^{(m)}(t,t)\mathbf{h}}},
\end{align*}
converge independently in distribution to Gaussian processes given the data in probability, similarly to the statement in Equation~(\ref{eq:asymp_distr_test_statistics}). 
Then, the assertions follow equivalently to the proof of Theorem~\ref{thm:distribution_test_statistics}.
Note that Proposition~\ref{thm:paired_permutation_clt} and the first assertion in Proposition~\ref{thm:uniform_consistency} prove also the permutation procedure in \textcite{smaga_note_2020}.
Firstly, a convergence statement equivalently to the assertion in Equation~(\ref{eq:diffclt}) is needed:
\begin{prop}\label{thm:paired_permutation_clt}
	Under \ref{ass:A5} and \ref{ass:A5}, it holds
	\begin{align}\label{eq:perm_diffclt}
		\left(\sqrt{n_1}(\hat\eta^{(c)}_{1,\tau}-\hat\eta^{(c)}_{2,\tau})(t)\right)_{t\in\mathcal{T}}\stackrel{d}{\longrightarrow}Y_{1-2}\sim \GP(0,\gamma_{1-2})
	\end{align}
	for $n_1\to\infty$ in $C(\mathcal{T})$, conditionally given the data in outer probability.
\end{prop}
\begin{proof}[Proof of Proposition~\ref{thm:paired_permutation_clt}]
	As written in Section~\ref{sec:eptheory}, because of Theorem~\ref{thm:clt}, the function class $\mathcal{F}$ is a Donsker class.
	With the dirac measure $\delta_A$, the empirical process $(\sqrt{n_1}(\hat\eta^{(c)}_{1}-\hat\eta^{(c)}_{2})(t))_{t\in\mathcal{T}}$ in Equation~(\ref{eq:diffclt}) can be rewritten as
	\begin{align*}
		\left(\sqrt{n_1}(\hat\eta^{(c)}_{1}-\hat\eta^{(c)}_{2})(t)\right)_{t\in\mathcal{T}}&=\left(\sqrt{n_1}\left(\frac{1}{n_1}\sum_{j=1}^{n_1}x^{(c)}_{1j}-\frac{1}{n_1}\sum_{j=1}^{n_1}x^{(c)}_{2j}\right)\right)_{t\in\mathcal{T}}\\
		&= \left(\sqrt{n_1}\frac{1}{n_1}\sum_{j=1}^{n_1}(x^{(c)}_{1j}-x^{(c)}_{2j})(t)\right)_{t\in\mathcal{T}} \\
		&=\left(\frac{1}{\sqrt{n_1}}\sum_{j=1}^{n_1}\left(\delta_{x^{(c)}_{1j}}-\delta_{x^{(c)}_{2j}}\right)f_t\right)_{t\in\mathcal{T}}\\
		&=\left(\frac{1}{\sqrt{n_1}}\sum_{j=1}^{n_1}\left(\mathbf{h}^\top\delta_{\mathbf{x}^{(c)}_j}-P\right)f_t\right)_{t\in\mathcal{T}}\\
		&=\left(\frac{1}{\sqrt{n_1}}\sum_{j=1}^{n_1}\left(\delta_{\mathbf{h}^\top\mathbf{x}^{(c)}_{j}}-P\right)f_t\right)_{t\in\mathcal{T}}\\
	\end{align*}
	for $f_t\in\mathcal{F}$ and $\mathbf{h}=(1,-1)^\top$ as in Section~\ref{sec:proof_distribution_test_statistics}.
	Therefore, the function class $\mathcal{F}_{1-2}:=\{f^-_t:C^2(\mathcal{T})\to\br|f^-_t(\mathbf{x}^{(c)}_j)=\mathbf{h}^\top\mathbf{x}^{(c)}_j(t),\,t\in\mathcal{T},\,j\in\{1,\dots,n_1\}\}$ is a Donsker class.
	Let $\xi_j$, $j\in\{1,\dots,n_1\}$\}, be some independently Rademacher distributed random variables, which means that it holds $\prob(\xi_j=1)=\prob(\xi_j=-1)=1/2$.
	Especially, $\xi_j$, $j\in\{1,\dots,n_1\}$, are independent of $\mathbf{x}^{(c)}$.
	Compare also Section 2.3.1 of \textcite{van_der_vaart_weak_2023} for more explanations on this choice.
	Then, it holds 
	\begin{align*}
		\left(\sqrt{n_1}(\hat\eta^{(c)}_{1,\tau}-\hat\eta^{(c)}_{2,\tau})(t)\right)_{t\in\mathcal{T}} &\stackrel{d}{=}\left(\sqrt{n_1}\frac{1}{n_1}\sum_{j=1}^{n_1}\xi_j(x^{(c)}_{1j}-x^{(c)}_{2j})(t)\right)_{t\in\mathcal{T}}\\
		&=\left(\frac{1}{\sqrt{n_1}}\sum_{j=1}^{n_1}\xi_j\left(\delta_{\mathbf{h}^\top\mathbf{x}^{(c)}_{j}}-P\right)f_t\right)_{t\in\mathcal{T}}
	\end{align*}
	for every $f_t\in\mathcal{F}$ and Theorem 2.9.6 in \textcite{van_der_vaart_weak_2023} can be applied.
	With Chapter 1.12 of \textcite{van_der_vaart_weak_2023}, it follows the assertion.
\end{proof}
For the incomplete data $\mathbf{x}^{(m)}$, the following convergence statement equivalently to the assertion in Equation~(\ref{eq:welchclt}) can be proven:

\begin{prop}\label{thm:unpaired_permutation_clt}
	Under \ref{ass:A1}, \ref{ass:A2}, and \ref{ass:A5}, it holds
	\begin{align}\label{eq:perm_welchclt}
		\left(\sqrt{\frac{n_2n_3}{n_2+n_3}}(\hat\eta^{(m)}_{1,\pi}-\hat\eta^{(m)}_{2,\pi})(t)\right)_{t\in\mathcal{T}}\stackrel{d}{\longrightarrow}G_H\sim\GP\left(0,\kappa_2\gamma_1+\kappa_3\gamma_2\right)
	\end{align}
	for $n_2+n_3\to\infty$ in $C(\mathcal{T})$ given the data in outer probability.
\end{prop}
\begin{proof}[Proof of Proposition~\ref{thm:unpaired_permutation_clt}]
	In the part of the data $\mathbf{x}^{(m)}$, where missing values are, the set of functions $\mathcal{F}^{(m)}=\{f_t^{(m)}:C^2(\mathcal{T})\to\br|f^{(m)}_t(\mathbf{x}_j^{(m)})=\mathbf{x}_j^{(m)},t\in\mathcal{T},j\in\{1,\dots,n_2+n_3\}\}$ is Donsker for both $P_1$ and $P_2$.
	Let 
	\begin{align*}
		\mathbb{P}_{1,n_2}=\frac{1}{n_2}\sum_{j=1}^{n_2}\delta_{x_{1j}}\;\text{and}\;\mathbb{P}_{1,n_2}=\frac{1}{n_3}\sum_{j=1}^{n_3}\delta_{x_{2j}},
	\end{align*}
	be the empirical measures for $\mathbf{x}_1^{(m)}$ and $\mathbf{x}_2^{(m)}$ and let
	\begin{align*}
		\mathbb{H}_{n_2+n_3}=\frac{1}{n_2+n_3}\sum_{j=1}^{n_2+n_3}\delta_{y_{j}}=\frac{n_2}{n_2+n_3}\mathbb{P}_{1,n_2}+\frac{n_3}{n_2+n_3}\mathbb{P}_{2,n_3},\,t\in\mathcal{T},
	\end{align*}
	be the pooled empirical measure.
	Then, we reformulate the empirical process in Equation~(\ref{eq:welchclt}) as $(\sqrt{n_2+n_3}(\hat\eta^{(m)}_1-\hat\eta^{(m)}_2)(t))_{t\in\mathcal{T}}=\sqrt{n_2+n_3}(\mathbb{P}_{1,n_2}-\mathbb{P}_{1,n_2})$.
	With \ref{ass:A4}, Theorem 3.8.1 in \textcite{van_der_vaart_weak_2023} can be applied on $\mathcal{F}^{(m)}$.
	From this, it follows with the empirical measures on the permuted data
	\begin{align*}
		\mathbb{\tilde P}_{1,n_2} =\frac{1}{n_2}\sum_{j=1}^{n_2}\delta_{y_{\pi(j)}}\;\text{and}\;\mathbb{\tilde P}_{1,n_2} =\frac{1}{n_3}\sum_{j=1}^{n_3}\delta_{y_{\pi(j)}}
	\end{align*}
	that 
	\begin{align*}
		\sqrt{\frac{n_2n_3}{n_2+n_3}}\left(\mathbb{\tilde P}_{1,n_2}- \mathbb{H}_{n_2+n_3}\right) &=\frac{\sqrt{n_3}}{\sqrt{n_2+n_3}}\sqrt{n_2}\left(\mathbb{\tilde P}_{1,n_2}- \mathbb{H}_{n_2+n_3}\right)
		\stackrel{d}{\longrightarrow}
		\sqrt{\kappa_3}G_H,\\
		\sqrt{\frac{n_2n_3}{n_2+n_3}}\left(\mathbb{\tilde P}_{2,n_3}- \mathbb{H}_{n_2+n_3}\right) &=\frac{\sqrt{n_2}}{\sqrt{n_2+n_3}}\sqrt{n_3}\left(\mathbb{\tilde P}_{2,n_3}- \mathbb{H}_{n_2+n_3}\right) \stackrel{d}{\longrightarrow}\sqrt{\kappa_2}G_H,	
	\end{align*}
	in $C(\mathcal{T})$ given the data in outer probability, where $G_H$ is a tight Brownian Bridge process corresponding to the measure $H=\kappa_2P_1+\kappa_3P_2$.
	From Theorem 1 in \textcite{munko_conditional_2026}, it follows that the latter two processes converge also together in probability.
	Consequently, it holds
	\begin{align*}
		&\left(\sqrt{\frac{n_2n_3}{n_2+n_3}}(\hat\eta^{(m)}_{1,\pi}-\hat\eta^{(m)}_{2,\pi})(t)\right)_{t\in\mathcal{T}}\\
		=\,&\sqrt{\frac{n_2n_3}{n_2+n_3}}\left(\mathbb{\tilde P}_{1,n_2}- \mathbb{\tilde P}_{1,n_2}\right)
		\stackrel{d}{\longrightarrow}
		\sqrt{\kappa_3}G_H
		+\sqrt{\kappa_2}G_H
		=G_H
	\end{align*}
	with $\kappa_2=1- \kappa_3$.
	The assertion follows with $\cov(G_H(t),G_h(s))=\kappa_2\gamma_1(s,t)+\kappa_3\gamma_2(s,t)$.
	
\end{proof}
Finally, it has to be argued that the permutation variance estimators $\{\hat\gamma_{1-2,\tau}^{(c)}(t,t):t\in\mathcal{T}\}$ $=\{\mathbf{h}^\top\boldsymbol{\hat\Gamma}^{(c)}_\tau(t,t) \mathbf{h}:t\in\mathcal{T}\}$ and 
$\{\hat\gamma_{\pi}^{(m)}(t,t):t\in\mathcal{T}\}=\{\mathbf{h}^\top\boldsymbol{\hat\Gamma}_{0,\pi}^{(m)}(t,t)\mathbf{h}:t\in\mathcal{T}\}$
are uniformly consistent. 
\begin{prop}\label{thm:uniform_consistency}
	Assume \ref{ass:A1},\ref{ass:A2}, \ref{ass:A3}, and \ref{ass:A5}.
	\begin{enumerate}
		\item The paired permutation variance estimator $\hat\gamma_{1-2,\tau}^{(c)}$ is uniformly consistent for $\gamma_{1-2}$, i.e.,
		\begin{align*}
			\sup_{t\in\mathcal{T}}\left\vert\hat\gamma_{1-2,\tau}^{(c)}(t,t)-\gamma_{1-2}(t,t)\right\vert\stackrel{p}{\longrightarrow}0
		\end{align*}
		given the data as $n_1\to\infty$.
		\item The unpaired permutation covariance estimator $\hat\gamma^{(m)}_{\pi}$ is uniformly consistent for $\gamma_{\pi}:=\var(G_H)=\kappa_2\gamma_1+\kappa_3\gamma_2$, i.e.,
		\begin{align*}
			\sup_{t\in\mathcal{T}}\left\vert \hat\gamma_{\pi}^{(m)}(t,t)-\gamma_{\pi}(t,t)\right\vert\stackrel{p}{\longrightarrow}0
		\end{align*}
		given the data as $n_2+n_3\to\infty$.
	\end{enumerate}
\end{prop}
\begin{proof}[Proof of Proposition~\ref{thm:uniform_consistency}]
	Regarding the first assertion, let $d^{(c)}_j(t)=x^{(c)}_{1j}(t)-x^{(c)}_{2j}(t)$ the pointwise differences and $d^{(c)}_{\tau(j)}(t)=x^{(c)}_{1\tau(j)}(t)-x^{(c)}_{2\tau(j)}(t)$ the pointwise permuted differences between the two groups for all $j\in\{1,\dots,n_1\}$ and $t\in\mathcal{T}$.
	The respective means are $\hat{\eta}_{1-2}:=\mathbf{h}^\top\boldsymbol{\hat{\eta}}$ or $\hat{\eta}_{1-2,\tau}:=\mathbf{h}^\top\boldsymbol{\hat{\eta}}_\tau$.
	Then, it holds for the paired permutation covariance estimator
	\begin{align}
		\nonumber \hat{\gamma}^{(c)}_{1-2,\tau}(t,t)&
		= \frac{1}{n_1-1} \sum_{j=1}^{n_1} \left(d_{j\tau(j)}^{(c)}(t)-\hat\eta_{1-2,\tau}^{(c)}(t)\right)^2 \\ 
		\nonumber&=\frac{1}{n_1-1} \sum_{j=1}^{n_1} \left(d_{\tau(j)}^{(c)}(t)\right)^2-\frac{n_1}{n_1-1}\left(\hat\eta_{1-2,\tau}^{(c)}(t)\right)^2\\ 
		\nonumber &\stackrel{d}{=}\frac{1}{n_1-1} \sum_{j=1}^{n_1} \left(\xi_jd_{j}^{(c)}(t)\right)^2-\frac{n_1}{n_1-1}\left(\hat\eta_{1-2,\tau}^{(c)}(t)\right)^2\\\label{eq:diffvariance_permutation}
		&=\frac{1}{n_1-1} \sum_{j=1}^{n_1} \left(d_{j}^{(c)}(t)\right)^2-\frac{n_1}{n_1-1}\left(\hat\eta_{1-2,\tau}^{(c)}(t)\right)^2.
	\end{align}
	Apart from that, as $\mathcal{F}$ is Donsker, it follows that $\mathcal{F}$ is also Glivenko-Cantelli \parencite[cf.][p. 130]{van_der_vaart_weak_2023}, i.e., it holds for $n_1\to\infty$
	\begin{align*}
		\sup_{t\in\mathcal{T}}\left\| \boldsymbol{\hat{\eta}}^{(c)}(t)-\boldsymbol{\eta}^{(c)}(t)\right\|_2 \stackrel{p}{\longrightarrow} 0,
	\end{align*}
	and consequently, from Theorem 3.8.1 in \textcite{van_der_vaart_weak_2023} it follows also
	\begin{align}\label{eq:gc_permutation}
		\sup_{t\in\mathcal{T}}\left\| \boldsymbol{\hat{\eta}}_\tau^{(c)}(t)-\boldsymbol{\hat{\eta}}^{(c)}(t)\right\|_2 \stackrel{p}{\longrightarrow} 0
	\end{align}
	given the data for $n_1\to\infty$.
	From this, we can conclude
	\begin{align*}
		&\sup_{t\in\mathcal{T}}\left\vert\hat{\gamma}^{(c)}_{1-2,\tau}(t,t)-\gamma_{1-2}^{(c)}(t,t)\right\vert\\
		\le& \sup_{t\in\mathcal{T}}\left\vert\hat{\gamma}^{(c)}_{1-2,\tau}(t,t)-\hat\gamma_{1-2}^{(c)}(t,t)\right\vert + \sup_{t\in\mathcal{T}}\left\vert\hat{\gamma}_{1-2}^{(c)}(t,t)-\gamma_{1-2}^{(c)}(t,t)\right\vert.
	\end{align*}
	As explained in Section~\ref{sec:methods}, the latter summand converges by Lemma 6 in \textcite{munko_multiple_2025} and by the Continuous Mapping Theorem in probability to $0$ if $n_1\to\infty$.
	Therefore, we only have to take care of the first summand and argue with Equation~(\ref{eq:diffvariance_permutation})
	\begin{align*}
		\sup_{t\in\mathcal{T}}\left\vert\hat{\gamma}^{(c)}_{1-2,\tau}(t,t)-\hat\gamma_{1-2}^{(c)}(t,t)\right\vert &= \sup_{t\in\mathcal{T}} \Bigg\vert\frac{1}{n_1-1} \sum_{j=1}^{n_1} \left(d_{j}^{(c)}(t)\right)^2-\frac{n_1}{n_1-1}\left(\hat\eta_{1-2,\tau}^{(c)}(t)\right)^2 \\
		&\quad\; - \quad \frac{1}{n_1-1} \sum_{j=1}^{n_1} \left(d_{j}^{(c)}(t)\right)^2-\frac{n_1}{n_1-1}\left(\hat\eta_{1-2}^{(c)}(t)\right)^2\Bigg\vert\\
		&= \frac{n_1}{n_1-1}\;\sup_{t\in\mathcal{T}}\left\vert \left(\hat\eta_{1-2,\tau}^{(c)}(t) - \hat\eta_{1-2}^{(c)}(t)\right)^2 \right\vert.
	\end{align*}
	From Equation~(\ref{eq:gc_permutation}), the Continuous Mapping Theorem and the definitions of the mean differences $\hat{\eta}_{1-2}=\mathbf{h}^\top\boldsymbol{\hat{\eta}}$ and $\hat{\eta}_{1-2,\tau}=\mathbf{h}^\top\boldsymbol{\hat{\eta}}_\tau$, it follows that the latter converges in probability to $0$, if $n_1\to\infty$, which is the first assertion.
	
	For the second assertion, we use some results of \textcite{janssen_studentized_1997}.
	He has proven the pointwise convergence of $\hat\gamma_{\pi}^{(m)}$ as part of Lemma 4.1, i.e., it holds
	\begin{align*}
		\hat\gamma_{\pi}^{(m)}(t,t)=\frac{n_3}{n_2+n_3}\hat\gamma_{1,\pi}(t,t)+\frac{n_2}{n_2+n_3}\hat\gamma_{2,\pi}(t,t)\stackrel{p}{\longrightarrow}\kappa_2\gamma_1(t,t)+\kappa_3\gamma_2(t,t)
	\end{align*}
	for every $t\in\mathcal{T}$ given the data.
	Therefore, by Theorem 2.1 of \textcite{newey_uniform_1991}, it remains to proof the stochastic equicontinuity of $\hat\gamma_{\pi}^{(m)}$. 
	To simplify the factors in the summands of $\hat\gamma_{\pi}^{(m)}$, define
	\begin{align*}
		c_j=
		\begin{cases}
			\frac{n_3}{(n_2+n_3)(n_3-1)},\quad j\in\{1,\dots,n_2\},\\
			\frac{n_2}{(n_2+n_3)(n_2-1)},\quad j\in\{1,\dots,n_3\},
		\end{cases}
	\end{align*}
	and compare \textcite{janssen_studentized_1997} for this approach.
	Then, by the definition of $\mathbf{v}_j(t)$, $j\in\{1,\dots,n\}$, $t\in\mathcal{T}$ in Assumption~\ref{thm:assumptions}, we can write for every $t\in\mathcal{T}$
	\begin{align}
		\hat\gamma_{\pi}^{(m)}=&\frac{n_3}{n_2+n_3}\left(\frac{1}{n_2-1}\sum_{j=1}^{n_2}\left(x_{1\pi(j)}(t)-\eta_{1,\pi}(t)\right)^2\right)\nonumber\\
		&+ \frac{n_2}{n_2+n_3} \left(\frac{1}{n_3-1} \sum_{j=1}^{n_3} \left(x_{2\pi(j)}(t)-\eta_{2,\pi}(t)\right)^2\right)\nonumber\\
		=&\frac{n_3}{n_2+n_3}\left(\frac{1}{n_2-1}\sum_{j=1}^{n_2}v_{1\pi(j)}(t)^2\right) + \frac{n_2}{n_2+n_3} \left(\frac{1}{n_3-1} \sum_{j=1}^{n_3} v_{1\pi(j)}(t)^2\right)\nonumber\\
		=&\sum_{j=1}^{n_2+n_3}c_j\left(v_{\pi(j)}^y(t)\right)^2,\label{eq:reformulation_permutation_variance}
	\end{align}
	where $v_{\pi(j)}^y(t)=y_{\pi(j)}(t)-\eta_{a_{\pi(j)},\pi(j)}(t)$ for every $t\in\mathcal{T}$ with 
	\begin{align*}
		a_{\pi(j)}=
		\begin{cases}
			1,\quad\pi(j)\in\{1,\dots,n_2\},\\
			2,\quad \pi(j)\in\{1,\dots,n_3\}.
		\end{cases}
	\end{align*}
	Then, to show the equicontinuity, we consider arbitrary $s,t\in\mathcal{T}$, choose $J$ as in \ref{ass:A2}, an arbitrary $\epsilon>0$, and apply Markov's Inequality:
	\begin{align}
		&P\left[\sup_{\zeta(s,t)<\delta}\left\vert \sum_{j=1}^{n_2+n_3}c_j\left(v_{\pi(j)}^y(s)\right)^2- \sum_{j=1}^{n_2+n_3}c_j\left(v_{\pi(j)}^y(t)\right)^2\right\vert>\epsilon\right]\nonumber\\
		\le\;& \E\left[\sup_{\zeta(s,t)<\delta}\left\vert \sum_{j=1}^{n_2+n_3}c_j\left(v_{\pi(j)}^y(s)\right)^2- \sum_{j=1}^{n_2+n_3}c_j\left(v_{\pi(j)}^y(t)\right)^2\right\vert^{J/2}\right]/\epsilon^{J/2}.\label{eq:markov}
	\end{align}
	Then, we have the following inequality:
	\begin{align}
		& \E\left[\left\vert \sum_{j=1}^{n_2+n_3}c_j\left(v_{\pi(j)}^y(s)\right)^2- \sum_{j=1}^{n_2+n_3}c_j\left(v_{\pi(j)}^y(t)\right)^2\right\vert^{J/2}\right]\nonumber\\
		\le\;& \E\left[\sum_{j=1}^{n_2+n_3}c_j^{J/2}\left\vert \left(v_{\pi(j)}^y(s)\right)^2- \left(v_{\pi(j)}^y(t)\right)^2\right\vert^{J/2}\right]\nonumber\\
		\le\;& 2^{J/2}\E\left[\left\vert \left(v_{\pi(1)}^y(s)\right)^2- \left(v_{\pi(1)}^y(t)\right)^2\right\vert^{J/2}\right]\nonumber\\
		=\;& 2^{J/2} \E\left[\left\vert \left(v_{\pi(1)}^y(s)\right)+ \left(v_{\pi(1)}^y(t)\right)\right\vert^{J/2}\,\left\vert \left(v_{\pi(1)}^y(s)\right)- \left(v_{\pi(1)}^y(t)\right)\right\vert^{J/2}\right]\nonumber\\
		\le\;& 2^{J/2}\sqrt{\E\left[\left\vert v_{\pi(1)}^y(s) + v_{\pi(1)}^y(t) \right\vert^J\right] \E\left[\left\vert v_{\pi(1)}^y(s)- v_{\pi(1)}^y(t)	\right\vert^J\right]}\nonumber\\
		\le\;& 2^{J-\frac{1}{2}}\sqrt{\E\left[\left\vert v_{\pi(1)}^y(s)\right\vert^J +\left\vert v_{\pi(1)}^y(t) \right\vert^J\right] \E\left[\left\vert v_{\pi(1)}^y(s)- v_{\pi(1)}^y(t)	\right\vert^J\right]}.\label{eq:equicontinuity}
	\end{align}
	Here, the first inequality follows from the triangle inequality, the second form \ref{ass:A1} and the fact that $n_i\ge2$, $i\in\{2,3\}$, then, the third binomial formula is used.
	The following inequality is deduced from the Cauchy-Schwarz inequality, and the last inequality is based on the binomial theorem of J-th order.
	Similarly to the approach in \textcite{janssen_studentized_1997} the expectation in the formula can be understood regarding the probability measure $P\otimes P^\pi$, where $P=(P_1,P_2)$ is the probability measure, that underlies the distribution of $\mathbf{x}$ and $P^\pi$ is the probability measure that underlies the random permutation $\pi$, which is independent from the data.
	Therefore, regarding the first factor under the root in Equation~(\ref{eq:equicontinuity}), we can argue for an arbitrary $t\in\mathcal{T}$:
	\begin{align*}
		\E\left[\left\vert v_{\pi(1)}^y(t) \right\vert^J\right]	=\;& \E_X\left[\E_\pi\left[\left\vert v_{\pi(1)}^y(t) \right\vert^J\bigg\vert X=x\right]\right]\\
		=\;& \E_X\left[\frac{1}{n_2+n_3}\sum_{j=1}^{n_2+n_3}\left\vert v_{j}^y(t) \right\vert^J\right] \\
		=\;&\frac{n_2}{n_2+n_3}\frac{1}{n_2}\sum_{j=1}^{n_2}\E_X\left[ \left\vert v_{11}(t) \right\vert^J\right]+ \frac{n_3}{n_2+n_3}\frac{1}{n_3}\sum_{j=1}^{n_3}\E_X\left[ \left\vert v_{21}(t) \right\vert^J\right]\\
		=\;&\frac{n_2}{n_2+n_3}\E_X\left[ \left\vert v_{11}(t) \right\vert^J\right]+ \frac{n_3}{n_2+n_3}\E_X\left[ \left\vert v_{21}(t) \right\vert^J\right]\\
		\le\;&\frac{n_2}{n_2+n_3}\E_X\left[\sup_{t\in\mathcal{T}} \left\vert v_{11}(t) \right\vert^J\right]+ \frac{n_3}{n_2+n_3}\E_X\left[ \sup_{t\in\mathcal{T}}\left\vert v_{21}(t) \right\vert^J\right] := D_J<\infty.
	\end{align*}
	The latter follows from \ref{ass:A2} and especially because of (R1) in  \textcite{munko_multiple_2025}.
	Equivalently, for the second factor under the root in Equation~(\ref{eq:equicontinuity}), we have
	\begin{align*}
		\E\left[\left\vert v_{\pi(j)}^y(s)- v_{\pi(j)}^y(t)	\right\vert^J\right]
		=\quad&\frac{n_2}{n_2+n_3}\E_X\left[ \left\vert v_{11}(s)-v_{11}(t) \right\vert^J\right]\\
		+& \frac{n_3}{n_2+n_3}\E_X\left[ \left\vert v_{21}(s) -v_{21}(t)\right\vert^J\right]
	\end{align*}
	and with \ref{ass:A5}, there exist some $C_i>0$ such that $\E_X[ \vert v_{i1}(s)-v_{i1}(t) \vert^J]\le\,C_i\zeta(s,t)^J$, $i\in\{1,2\}$.
	Applying this and $D_J<\infty$ on Equation~(\ref{eq:equicontinuity}),
	it holds
	\begin{align*}
		\E\left[\left\vert \sum_{j=1}^{n_2+n_3}c_j\left(v_{\pi(j)}^y(s)\right)^2- \sum_{j=1}^{n_2+n_3}c_j\left(v_{\pi(j)}^y(t)\right)^2\right\vert^{J/2}\right]
		\le \tilde{C}\zeta(s,t)^{J/2}
	\end{align*}
	with $\tilde{C}:=2^{J}\sqrt{D_J(C_1+C_2)}>0$, where $\tilde{C}$ depends only on $J$, $D_J$, and $C_i$, $i\in\{1,2\}$, and not on $s,t\in\mathcal{T}$.
	Consequently, the assumptions of Theorem 2.2.4. in \textcite{van_der_vaart_weak_2023} are satisfied.
	Therefore, with $\psi:\br^{\ge 0}\to\br$, $\psi(x)=x^{J/2}$ and $d=\zeta$, there exists a constant $K$ depending only on $J$, $D_J$, and $C_i$, $i\in\{1,2\}$, such that
	\begin{align*}
		&\E\left[\sup_{\zeta(s,t)<\delta}\left\vert \sum_{j=1}^{n_2+n_3}c_j\left(v_{\pi(j)}^y(s)\right)^2- \sum_{j=1}^{n_2+n_3}c_j\left(v_{\pi(j)}^y(t)\right)^2\right\vert^{J/2}\right]\\
		\le\;& K \left(\int_{0}^{\nu}D(\epsilon, \zeta)^{2/J}d\epsilon+\delta D(\nu,\zeta)^{4/J}\right)^{J/2}
	\end{align*}
	holds for all $\nu,\delta>0$.
	Because of Equation~(\ref{eq:packingnumber}) in \ref{ass:A5} and since $\nu$ can be chosen arbitrarily small, it follows that
	\begin{align*}
		\lim_{\delta \searrow0}\limsup_{n_2+n_3\to\infty}\E\left[\sup_{\zeta(s,t)<\delta}\left\vert \sum_{j=1}^{n_2+n_3}c_j\left(v_{\pi(j)}^y(s)\right)^2- \sum_{j=1}^{n_2+n_3}c_j\left(v_{\pi(j)}^y(t)\right)^2\right\vert^{J/2}\right]=0
	\end{align*}
	With Equation~(\ref{eq:markov}), we deduce
	\begin{align*}
		\lim_{\delta \searrow0}\limsup_{n_2+n_3\to\infty}P\left[\sup_{\zeta(s,t)<\delta}\left\vert \sum_{j=1}^{n_2+n_3}c_j\left(v_{\pi(j)}^y(s)\right)^2- \sum_{j=1}^{n_2+n_3}c_j\left(v_{\pi(j)}^y(t)\right)^2\right\vert^{J/2}>\epsilon\right]=0,
	\end{align*}
	which shows the stochastic equicontinuity of $\hat\gamma_\pi^{(m)}$ by Equation~(\ref{eq:reformulation_permutation_variance}).
	It follows the assertion.
\end{proof}

Equivalently to Equation~(\ref{eq:asymp_distr_test_statistics}) in the proof of Theorem~\ref{thm:distribution_test_statistics}, from the Propositions~\ref{thm:paired_permutation_clt},~\ref{thm:unpaired_permutation_clt} and \ref{thm:uniform_consistency}, it can be concluded with Slutzky`s Theorem that
\begin{align*}
	\left(T_{n_1}(\mathbf{x}_\tau^{(c)},\boldsymbol{\hat\Gamma}_\tau^{(c)})(t),\,T_{n_2+n_3}(\mathbf{x}_\pi^{(m)}, \boldsymbol{\hat \Gamma}_{0,\pi}^{(m)})(t) \right)_{t\in\mathcal{T}}\stackrel{d}{\longrightarrow}(Z_1,Z_{2,\pi})
\end{align*}
in $C(\mathcal{T})^2$ conditionally, given the data in probability with
\begin{align*}
	\cov(Z_{2,\pi}(t),Z_{2,\pi}(s))=\,&\cov\left(\frac{G_H(t)}{\sqrt{\kappa_2\gamma_1(t,t)+\kappa_3\gamma_2(t,t)}},\, \frac{G_H(s)}{\sqrt{\kappa_2\gamma_1(s,s)+\kappa_3\gamma_2(s,s)}}  \right)\\
	=\,& \frac{\cov\left(G_H(t),G_h(s)\right)}{\sqrt{\kappa_2\gamma_1(t,t)+\kappa_3\gamma_2(t,t)}\sqrt{\kappa_2\gamma_1(s,s)+\kappa_3\gamma_2(s,s)}}\\
	=\,& \frac{\kappa_2\gamma_1(s,t)+\kappa_3\gamma_2(s,t)}{\sqrt{\kappa_2\gamma_1(t,t)+\kappa_3\gamma_2(t,t)}\sqrt{\kappa_2\gamma_1(s,s)+\kappa_3\gamma_2(s,s)}}.
\end{align*}
and $\var(Z_{2,\pi}(t))=1$ for all $s,t\in\mathcal{T}$.
Equivalently to the proof of Theorem~\ref{thm:distribution_test_statistics}, the assertion follows.

\subsection{Proof of Theorem~\ref{thm:distribution_boot_test_statistic}} 
The same resampling procedure is used in \textcite{munko_multiple_2025} to infer a functional \textsc{manova} model without missing values.
That is why we can use several results of \textcite{munko_multiple_2025} to proof the theorem.

The following statement is a special case of Lemma 7 in \textcite{munko_multiple_2025}:
we have convergence in distribution
\begin{align}\label{eq:boot_diffclt}
	\left(\sqrt{n_1}(\hat\eta^{(c)}_{1,\mathcal{P}}-\hat\eta^{(c)}_{2,\mathcal{P}})(t)\right)_{t\in\mathcal{T}}\stackrel{d}{\longrightarrow}Y_{1-2}\sim \GP(0,\gamma_{1-2})
\end{align}
for $n_1\to\infty$ in $C(\mathcal{T})$, conditionally given the data in outer probability, where the vector $\boldsymbol{\hat{\eta}}^{(c)}_\mathcal{P}=(\hat\eta^{(c)}_{1,\mathcal{P}},\hat\eta_{2,\mathcal{P}}^{(c)})^\top$ contains the group-wise means $\hat\eta^{(c)}_{1,\mathcal{P}}=1/n_1\sum_{j=1}^{n_1}x^{(c)}_{1j,\mathcal{P}}$ and $\hat\eta^{(c)}_{2,\mathcal{P}}=1/n_1\sum_{j=1}^{n_1}x^{(c)}_{2j,\mathcal{P}}$ on $\mathbf{x}^{(c)}_\mathcal{P}$.
From Equation~(29) in Lemma~7 of \textcite{munko_multiple_2025}, with \ref{ass:A4} and the Continuous Mapping Theorem it follows that
\begin{align}\label{eq:boot_welchclt}
	\left(\sqrt{\frac{n_2n_3}{n_2+n_3}}(\hat\eta^{(m)}_{1,\mathcal{P}}-\hat\eta^{(m)}_{2,\mathcal{P}})(t)\right)_{t\in\mathcal{T}}\stackrel{d}{\longrightarrow}\sqrt{\kappa_3}\,Y_1-\sqrt{\kappa_2}\,Y_2\sim\GP\left(0,\kappa_3\gamma_1+\kappa_2\gamma_2\right)
\end{align}
for $\min\{n_2,n_3\}\to\infty$ in $C(\mathcal{T})$ conditionally given the data in outer probability, where the vector $\boldsymbol{\hat{\eta}}^{(m)}_\mathcal{P} = (\hat\eta^{(m)}_{1,\mathcal{P}}, \hat\eta^{(m)}_{2,\mathcal{P}})^\top$ includes the group-wise means on $\mathbf{x}^{(m)}_\mathcal{P}$.
Lemma~8 in \textcite{munko_multiple_2025} ensures the uniform consistency of the parametric bootstrap covariance estimator $\boldsymbol{\hat{\Gamma}}_{\mathcal{P}}$.
As all considered variances are continuous functions of this estimator or of its diagonal elements $\hat\gamma_{1,\mathcal{P}}$ and $\hat\gamma_{2,\mathcal{P}}$, this ensures the uniform consistency of the variance estimator of the paired part $\hat\gamma_{1-2,\mathcal{P}}$ and the variance estimator of the unpaired part $n_3/(n_2+n_3)\hat\gamma_{1,\mathcal{P}}+n_2/(n_2+n_3)\hat\gamma_{2,\mathcal{P}}$.
From this and Equations~(\ref{eq:boot_diffclt})-(\ref{eq:boot_welchclt}) it can be concluded similar to the unconditional case that by Slutzky
\begin{align*}
	\left(T_{n_1}(\mathbf{x}_\mathcal{P}^{(c)},\boldsymbol{\hat\Gamma}_\mathcal{P}^{(c)})(t),\,T_{n_2+n_3}(\mathbf{x}_\mathcal{P}^{(m)}, \boldsymbol{\hat \Gamma}_{0,\mathcal{P}}^{(m)})(t) \right)_{t\in\mathcal{T}}\stackrel{d}{\longrightarrow}(Z_1,Z_2)
\end{align*}
holds in $C(\mathcal{T})^2$ conditionally, given the data in probability.
The argument is equivalent to Equation~(\ref{eq:asymp_distr_test_statistics}) in the proof of Theorem~\ref{thm:distribution_test_statistics}. 
Similar to the proof of Theorem~\ref{thm:distribution_test_statistics}, the assertion follows.

\end{document}